\documentclass[twocolumn]{emulateapj}
\usepackage{amsmath}
\usepackage{graphicx}
\usepackage{natbib}  
\newcommand{\kpc}{\,{\rm kpc}}

\newcommand{\kms}{\,km\,s$^{-1}$}  
\newcommand{\myr}{\,$M_{\sun}\,{\rm yr}^{-1}$}

\newcommand{\mo}{\,$M_{\sun}$}
\newcommand{\lo}{\,$L_{\sun}$}
\newcommand{\cmt}{\,cm$^{-3}$}

\newcommand{\cmd}{\,cm$^{-2}$}

\newcommand{\es}{$\rm\,erg\,s^{-1}$}

\newcommand{\ecsa}{$\rm\,erg\,cm^{-2}\,s^{-1}\,\AA^{-1}$}
\slugcomment{Accepted version May 01, 2019}

\shorttitle{Multiwavelength modeling the SED of nova V339~Del}
\shortauthors{Skopal}

\begin{document}

%
%
%
\title{Multiwavelength modeling the SED of nova V339~Del: 
Stopping the wind \\
and long-lasting super-Eddington luminosity with dust emission }

\author{Augustin Skopal\altaffilmark{1}}
\affil{Astronomical Institute, Slovak Academy of Sciences,
       059\,60 Tatransk\'{a} Lomnica, The Slovak Republic}
\altaffiltext{1}{E-mail: skopal@ta3.sk}

\begin{abstract}
During the classical nova outburst, the radiation generated by 
the nuclear burning of hydrogen in the surface layer of a white 
dwarf (WD) is reprocessed by the outer material into different 
forms at softer energies, which distribution in the spectrum 
depends on the nova age. 
Using the method of multiwavelength modeling the SED we 
determined physical parameters of the stellar, nebular and 
dust component of radiation isolated from the spectrum of 
the classical nova V339~Del from day 35 to day 636 after 
its explosion. 
The transition from the iron-curtain phase to the super-soft 
source phase (days 35--72), when the optical brightness dropped 
by 3--4\,mag, the absorbing column density fell by its 
circumstellar component from $\sim 1\times 10^{23}$ to 
$\sim 1\times 10^{21}$\cmd, and the emission measure decreased 
from $\sim 2\times 10^{62}$ to $\sim 8.5\times 10^{60}$\cmt, 
was caused by stopping-down the mass-loss from the WD. 
The day 35 model SED indicated an oblate shape of the WD 
pseudophotosphere and the presence of the dust located 
in a slow equatorially concentrated outflow. The dust emission 
peaked around day 59. Its co-existence with the strong super-soft 
X-ray source in the day 100 model SED constrained the presence 
of the disk-like outflow, where the dust can spend a long time. 
Both the models SED revealed a super-Eddington luminosity of 
the burning WD at a level of 
$1-2\times 10^{39}\,(d/4.5\kpc)^2$\es, 
lasting from $\sim$day 2 to at least day 100. 
%
%
\end{abstract}
\maketitle
%
%
\section{Introduction}
\label{s:intro}
The nova phenomenon results from a thermonuclear runaway (TNR) 
on the surface of a white dwarf (WD) accreting hydrogen-rich 
material from its companion in a binary system 
\citep[e.g.][ for a review]{bode08,starrfield+16}. 
The TNR event significantly increases the luminosity and 
ejects material of $10^{-3} - 10^{-7}$\mo\ at velocities 
$\gtrsim 10^3$\kms\ \citep[e.g.,][]{gall+star78,k+h94}. 
The energy released near the WD surface is thus reprocessed 
through the outer material giving rise to the spectral energy 
distribution (SED) that changes across the electromagnetic 
spectrum with the nova age. 
In this work we introduce modeling the SED of a classical 
nova V339~Del at critical dates of its evolution to obtain 
new information about the nova explosion. 

Classical nova V339~Del (Nova Delphini 2013 = PNV J20233073+2046041) 
was discovered by Koichi Itagaki on 2013 August 14.584 UT 
at a visual magnitude of $\sim$6.8 \citep[][]{nakano13}. 
Its progenitor was identified by \cite{denisenko+13} as the blue 
star USNO-B1.0 1107-0509795 ($B\sim 17.2-17.4$, $R\sim 17.4-17.7$). 
\cite{m+h13} found the progenitor also on the Asiago 1979-82 
plates and within observations taken by the APASS 
survey\footnote{http://www.aavso.org/apass} in April 2012, 
whereas \cite{deacon+14} documented its pre-outburst variability. 
%
%
After $\sim$1.85 days of its discovery, the nova peaked at 
$V\sim 4.43$ on August 16.45 UT \citep[][]{munari+13b}, and 
became to be an attractive target also for amateurs astronomers. 
As a result, a large amount of observations with a high cadence 
in the optical, has been performed from $\gamma$-rays 
\citep[][]{acker+14,ahnen+15} to the radio/mm--2\,cm wavelengths 
\citep[][]{chomiuk+13,anderson+13}. 
A short review of V339~Del was presented by \cite{munari+15}, 
\cite{shore+16}, \cite{evans+17} and \cite{chochol+17}. 
Figure~\ref{fig:lc} shows the optical light curve (LC) of the nova 
from day $\sim$1 after its explosion (see below) to day 
$\sim$1800. 
Main characteristics of V339~Del can be summarized as follows. 

{\sc Reddening:} 
Based on the equivalent width of Na\,{\small I}\,$\lambda$5890, 
\cite{munari+13a} determined the interstellar reddening to 
V339~Del to be $E_{\rm B-V}$ = 0.18, which agrees with the 
\cite{schlegel+98} extinction maps as found by \cite{burlak+15}. 
This value has been used in most papers published to date. 
%
%
\begin{figure*}
\begin{center}
\resizebox{\hsize}{!}{\includegraphics[angle=-90]{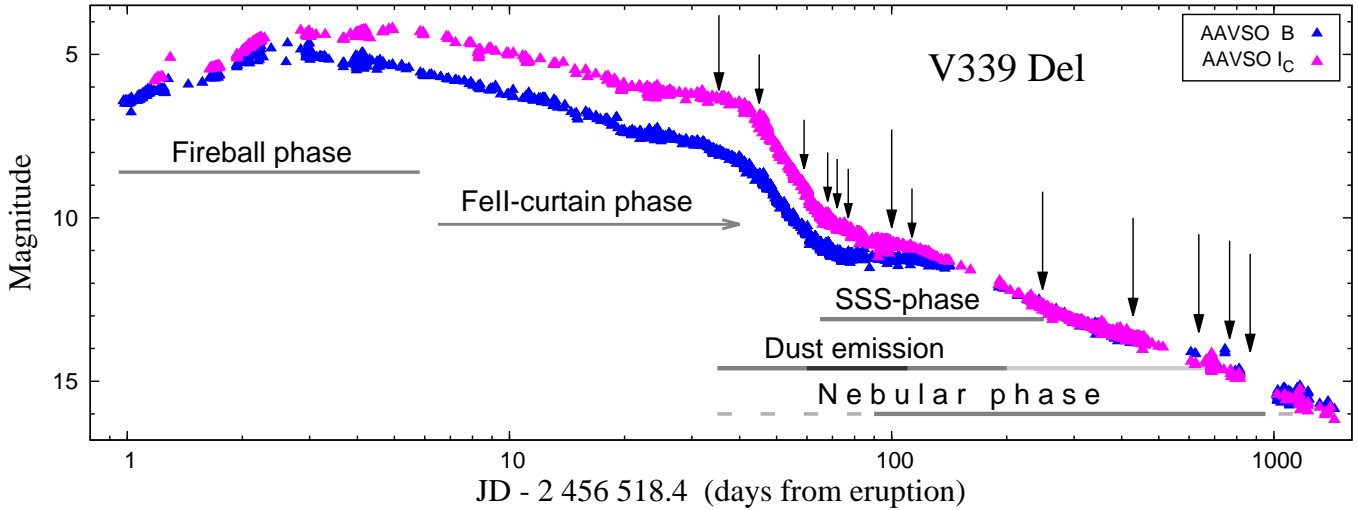}}
\end{center}
\caption{
The $B$ and $I_{\rm C}$ LCs of V339~Del from day 1 to day 
$\sim$1500 from the nova eruption as collected in the 
\textsl{AAVSO} International Database. Long and short arrows 
indicate times of the \textsl{HST/STIS} and optical/near-IR 
observations, respectively (Table~\ref{tab:obs}). 
}
\label{fig:lc}
\end{figure*}

{\sc Distance:} 
Using the maximum-magnitude-rate-of-decline relation, 
the distance to the nova was estimated to 2.7--3.5\kpc\ 
\citep[][]{taranova+14,chochol+14,burlak+15}. 
On the basis of near-IR interferometry, \cite{schaefer+14} 
determined the pre-maximum expansion rate of the nova fireball, 
which corresponds to a distance of $4.54\pm 0.59$\kpc. 
In a similar way, but using photometry of the expanding fireball, 
\cite{gehrz+15} determined the distance of V339 Del to 
$4.5\pm 0.8$\kpc\ and the day zero (= the nova explosion) on 
JD 2~456\,518.4 (August 13.9, 2013). The distance of $4-4.5$\kpc\ 
was supported by \cite{shore+16} by comparing the UV spectrum of 
the CO nova OS~And 1986 and V339~Del taken at a comparable stage 
of their evolution. 
Using the distance-extinction relation for the independent  
measurements of reddening, \cite{ozdonmez+16} and \cite{ozdonmez+18} 
included V339~Del to novae, for which the distance could 
not be estimated. 
Finally, the Gaia data release 2 catalog presents the distance 
of 2130$^{+2250}_{-400}$\,pc, which was classified by 
\cite{schaefer18} as untrustworthy, because having the parallax 
error bar $>30$\,\%. 

{\sc Temperature:}
During the fireball stage (August 14.8--19.9, 2013; day 0.9--6.0), 
when the maximum of the nova radiation was within the optical, 
\cite{sk+14} determined the effective temperature of the WD 
pseudophotosphere in the range of 6000--12000\,K on the basis 
of models SED. 
Comparing photometric $UBVRIJHKLM$ flux-points from the optical 
maximum (August 15.94 and 16.86, 2013; day 2 and 3) to the 
spectrum of normal supergiants, \cite{taranova+14} estimated the 
temperature of the nova to 13600 and 9400\,K, respectively. 
%
During the super-soft source (SSS) phase, on 
November 09.75, 2013 (day 88), 
\cite{nelson+13} estimated a photospheric temperature of 
the WD to 27\,eV ($\sim $310000\,K) on the basis of the 
first high resolution X-ray spectrum of the nova obtained 
by the \textsl{Chandra} observatory. 
On November 21, 2013 (day 100), \cite{ness+13} estimated 
the temperature of the SSS to $\sim $30\,eV ($\sim$350000\,K) 
using the spectrum obtained with \textsl{XMM-Newton}. 

{\sc Luminosity:} 
Around the optical maximum, \cite{taranova+14}, \cite{sk+14} 
and \cite{gehrz+15} estimated the luminosity of the nova to 
$L_{\rm WD} \approx 2.5\times 10^5\,(d/3\,{\rm kpc})^2$, 
$(2.2\pm 0.2)\times 10^5\,(d/3\,{\rm kpc})^2$ and 
$\sim 8.3\times 10^5\,(d/4.5\,{\rm kpc})^2$\lo, respectively, 
i.e., a factor of $\sim$10 above the Eddington value. 
Using the \textsl{HST/STIS} and optical observations on 
day 35 and 100, \cite{shore+16} determined the 1200--7400\,\AA\ 
luminosity to $(5\pm 0.5)\times 10^4$\lo, which is still at 
or above the Eddington value for a Chandrasekhar mass WD. 

{\sc Dust formation:} 
On the basis of near-IR photometry, \cite{taranova+14} indicated 
a dust emission around one month after the optical maximum. 
On September 21 (day 38.9) and October 11, 2013 (day 58.8), 
they estimated its color temperature to $\approx$1500 and 
$\approx$1200\,K, luminosity of $\sim$4$\times 10^3$ and 
$\sim$1.2$\times 10^4\,(d/3\,{\rm kpc})^2$\lo\ and its mass to 
$\sim$1.6$\times 10^{24}$ and $\sim$1.1$\times 10^{25}$\,g, 
respectively. 
Using the stratospheric observatory \textsl{SOFIA} and ground-based 
infrared observations, \cite{gehrz+15} revealed the presence of 
a dust emission with a blackbody temperature of $(850\pm 40)$\,K, 
luminosity of $\sim$1.7$\times 10^4\,(d/4.5\,{\rm kpc})^2$\lo\ and 
a mass of $(1.2\pm 0.4)\times 10^{-7}$\mo\ on November 24.04, 
2013 (day 102.14). 
An extensive description of the dust development in V339~Del 
was presented by \cite{evans+17}, who found that dust formation 
commenced on day $\sim$34.75 at condensation temperature of 
1480$\pm 20$\,K, consistent with graphitic carbon. They determined 
the rise and fall of the mass of the dust with a maximum at 
day $\sim$100 and the last detection on day 636. 
\cite{evans+17} and \cite{shore+18} considered a possibility 
of a dust extinction effect in the optical and UV LC, although 
a strong dust emission in the near-IR was unambiguously 
indicated (see Sect.~\ref{sss:dust} here). 

{\sc The ejecta mass:} 
Based on the emission measure (\textsl{EM}) from models SED, 
\cite{sk+14} determined the ejecta mass to 
$\sim$1$\times 10^{-5}$ -- $\sim$4.6$\times 10^{-4}$\mo\ 
between day 8 and 38, for the volume filling factor $f = 1$. 
Using \textsl{EM} and $F_{\rm H\beta}$ flux measured over the 
period 253--352 days, \cite{tar+sk16} estimated the ejecta mass 
to $\sim$7$\times 10^{-5}$\mo\ for $f = 0.28$.  
\cite{gehrz+15} obtained $\sim$7.5$\times10^{-5}$\mo\ by estimating 
the cutoff wavelength during the free--free emission phase, and 
\cite{shore+16} derived a range of the ejecta mass to 
$(2-3)\times 10^{-5}$\mo\ for $f = 0.1$. 

{\sc Veiling the WD photosphere:} 
It is best documented by the X-ray emission, which is effectively 
attenuated by the neutral atoms of hydrogen. 
X-ray radiation from V339~Del was monitored by the {\em Swift} 
X-ray telescope with the first detection at harder energies 
(1--10\,keV) on September 19 (day 37), whereas a soft component 
was detected on October 13, 2013 (day 61), indicating the start 
of visibility of the WD photosphere \citep[][]{page+13}. 
Using the data from September 19 to 26, 2013 (day 37 to 44) 
\cite{pagebead13} determined the absorbing hydrogen column 
density, $N_{\rm H} = 4.9^{+7.5}_{-3.2}\times 10^{22}$\cmd. 
A decrease of $N_{\rm H}\sim 5\times 10^{22}$ around day 45 
to $\sim 1.8\times 10^{22}$\,\cmd\ around day 65 was reported 
by \cite{osborne+13}. 
On November 09.75, 2013 (day 88), \cite{gatuzz+18} determined 
$N_{\rm H}\sim 1.23\times 10^{21}$\,\cmd\ 
using the \textsl{Chandra} X-ray spectrum. 
A summary of the X-ray evolution is given by \cite{shore+16}. 
%
%
\begin{table*}
\caption[]{Log of observations}
\begin{center}
\begin{tabular}{lllcc}
\hline
\hline
 ~~Age$^a$ & ~~~~~Date UT  & Julian date & Region & Observatory/Ref.  \\
JD--JD$_0$ &    YYYY-MM-DD & JD~2\,456...&        &                   \\
\hline
%
~~34.817 & 2013-09-17.717 & 553.217 & $JHKLM$       & CrAO$^b$ \\
~~35     & 2013-09-17.5   & 553     &  $UBV$        & CrAO$^b$ \\
~~35.319 & 2013-09-18.219 & 553.719 &115--289\,nm   & \textsl{HST/STIS} \\
~~35.3   & 2013-09-18.2   & 553.7   &  $BVR_C$      & \textsl{AAVSO}$^c$ \\
~~35.7   & 2013-09-18.6   & 554.1   &360--965\,nm   & FKO$^d$ \\
~~36.82  & 2013-09-19.72  & 555.22  &1.15--2.36\,$\mu$m & Mt~Abu$^e$ \\
\hline
%
~~44.616 & 2013-09-27.516 & 563.016 &365--911\,nm   & FKO$^d$ \\
~~44.869 & 2013-09-27.769 & 563.269 & $JHKLM$       & CrAO$^b$ \\
~~45     & 2013-09-27.5   & 563     &  $UBV$        & CrAO$^b$ \\
~~45.194 & 2013-09-28.094 & 563.594 & $BVR_CI_C$    & $^n$ \\
\hline
%
~~58.788 & 2013-10-11.688 & 577.188 & $JHKLM$       & CrAO$^b$ \\
~~59     & 2013-10-11.5   & 577     &  $UBV$        & CrAO$^b$ \\
~~59.572 & 2013-19-12.472 & 577.972 & $BVR_CI_C$    & $^n$ \\
~~59.637 & 2013-10-12.537 & 578.037 &365--930\,nm   & FKO$^d$ \\
\hline
%
~~67.832 & 2013-10-20.732 & 586.232 & $JHKLM$       & CrAO$^b$ \\
~~68     & 2013-10-20.5   & 586     &  $UBV$        & CrAO$^b$ \\
~~68.318 & 2013-10-21.218 & 586.718 & $BVR_CI_C$    & $^n$ \\
~~68.582 & 2013-10-21.482 & 586.982 &365--930\,nm   & FKO$^d$ \\
\hline
%
~~71.817 & 2013-10-24.717 & 590.217 & $JHKLM$       & CrAO$^b$ \\
~~71.875 & 2013-10-24.775 & 590.275 &363--720\,nm   & \textsl{ARAS}$^o$ \\
~~72     & 2013-10-24.5   & 590     &  $UBV$        & CrAO$^b$ \\
~~72.236 & 2013-10-25.014 & 590.636 & $BVR_CI_C$    & $^n$ \\
~~74.542 & 2013-10-27.442 & 592.942 &370--900\,nm   & FKO$^d$ \\
\hline
%
~~75.837 & 2013-10-28.737 & 594.237 & $BVR_CI_C$   & $^n$ \\
~~76.770 & 2013-10-29.670 & 595.170 & $JHKLM$      & CrAO$^b$ \\
~~76.81  & 2013-10-29.71  & 595.21  &  $UBV$       & CrAO$^b$ \\
~~77.534 & 2013-10-30.434 & 595.934 &371--900\,nm  & FKO$^d$ \\
~~77.922 & 2013-10-30.822 & 596.322 &375--740\,nm  & \textsl{ARAS}$^o$ \\
\hline
%
~~99.845& 2013-11-21.745 & 618.245 &2.3--3.8\,nm   & \textsl{XMM-Newton} \\
~~99.957& 2013-11-21.857 & 618.357 &379--736\,nm   & \textsl{ARAS}$^o$ \\
100     & 2013-11-21.5   & 618     &   $UBV$       & CrAO$^b$ \\
100.0   & 2013-11-21.9   & 618.4   & $BVR_CI_C$    & \textsl{AAVSO}$^h$ \\
100.029 & 2013-11-21.929 & 618.429 &116--307\,nm   & \textsl{HST/STIS} \\
100.510 & 2013-11-22.410 & 618.910 &365--911\,nm   & FKO$^d$ \\
101.66  & 2013-11-23.56  & 620.06  &   $JHK$       & Mt~Abu$^i$ \\
102.14  & 2013-11-24.04  & 620.54  &  $KLMNN'$     & OBO$^i$ \\
102.66  & 2013-11-24.56  & 621.06  &1.14--11.6\,$\mu$m & Mt~Abu$^e$ \\
\hline
%
111.835 & 2013-12-03.735 & 630.235 &375--740\,nm   & \textsl{ARAS}$^o$ \\
112.759 & 2013-12-04.659 & 631.159 & $JHKLM$       & CrAO$^b$ \\
113.1   & 2013-12-05.0   & 631.5   & $BVR_CI_C$    & \textsl{AAVSO}$^p$ \\
113     & 2013-12-04.5   & 631     &  $UBV$        & CrAO$^b$ \\
113.852 & 2013-12-05.752 & 632.252 &662--885\,nm   & \textsl{ARAS}$^o$ \\
\hline

248.17  & 2014-04-19.07  & 766.57  &115--570\,nm   & \textsl{HST/STIS} \\
248.47  & 2014-04-19.3   & 766.87  & $BVR_CI_C$    & \textsl{AAVSO}$^j$ \\
253.871 & 2014-04-24.771 & 772.271 &400--820\,nm   & FKO$^d$ \\
255.075 & 2014-04-25.975 & 773.475 &380--750\,nm   & CrAO$^k$ \\
\hline
428.116 & 2014-10-16.016 & 946.516 &115--307\,nm   & \textsl{HST/STIS} \\
428.1   & 2014-10-16.0   & 946.5   & $BVR_CI_C$    & \textsl{AAVSO}$^l$ \\
393.622 & 2014-09-11.522 & 912.022 &400--820\,nm   & FKO$^d$ \\
457.536 & 2014-11-14.436 & 975.936 &430--730\,nm   & FKO$^d$ \\
\hline
636.076 & 2015-05-11.976 &1154.476 &115--307\,nm   & \textsl{HST/STIS} \\
636.1   & 2015-05-12.0   &1154.5   & $BVR_CI_C$    & \textsl{AAVSO}$^m$ \\
644.851 & 2015-05-20.751 &1163.251 &400--743\,nm   & FKO$^d$ \\
\hline
766.198 & 2015-09-19.098 &1284.598 &115--171\,nm   & \textsl{HST/STIS} \\
\hline
867.126 & 2015-12-29.026 &1385.526 &115--171\,nm   & \textsl{HST/STIS} \\
\hline
\end{tabular}
\end{center}
$^a$JD$_0$ = 2\,456\,518.4 (August 13.9, 2013) is the date of 
                nova explosion, 
$^b$\cite{burlak+15}, 
$^c$$B=8.0,~V=7.98,~R_C=6.44$, 
$^d$Fujii Kurosaki Observatory \citep[see][]{sk+14}, 
$^e$representative fluxes from Fig.~7 of \cite{evans+17}, 
$^h$$B=11.24, V=11.20, R_C=10.36, I_C=10.85$, 
$^i$see Table~4 of \cite{gehrz+15}, 
$^j$$B=12.67, V=11.88, R_C=12.02, I_C=12.67$, 
$^k$\cite{tar+sk16},
$^l$$B=13.72, V=12.62, R_C=12.96, I_C=13.69$,
$^m$$B=14.22, V=13.25, R_C=13.45, I_C=14.47$,
$^n$\cite{munari+13b},
$^o$http://www.astrosurf.com/aras/Aras$\_ $DataBase/Novae/Nova-Del-2013.htm,
$^p$$B=11.31, V=11.24, R_C=10.43, I_C=10.90$. 
\label{tab:obs}
\end{table*}

{\sc Geometry of the nova:} 
Using the near-IR interferometry, \cite{schaefer+14} indicated 
a prolate structure of V339~Del already 2 days after the 
eruption. 
Modeling the optical/near-IR SED during first 40 days of 
the nova evolution, \cite{sk+14} suggested a biconical 
ionization structure of the ejecta with a disk-like H\,{\small I} 
region encompassing the WD at the orbital plane. 
During the nebular phase, \cite{tar+sk16} concluded that the 
ejected material has a disk-polar structure with the orbital
inclination of $\sim 65^{\circ}$. Values of $i = 55-35^{\circ}$ 
were suggested by \cite{shore+16}. 
Using high-resolution spectropolarimetric observations, 
\cite{kawakita+19} indicated an expanding equatorial torus 
surrounding the nova photosphere, which changed to a bipolar 
geometry during a few days of the nova age. 

These results were achieved by analyzing the data covering 
the X-ray, ultraviolet, optical and near-infrared part of the 
nova spectrum, in most cases, separately. However, the analysis 
of multiwavelength observations covering a wide energy range 
obtained simultaneously at different nova ages can provide 
a better understanding of the nova evolution. 

Accordingly, we reconstruct and model the SED of V339~Del 
at days when a large fraction of its spectrum is covered by 
nearly-simultaneous observations taken from day 35 
to day 636. 
In Sect.~\ref{s:obs} we summarize the used observations, 
while Sect.~\ref{s:analysis} describes our analysis and 
presents the results. Their discussion and summary are found 
in Sects.~\ref{s:dis} and \ref{s:sum}, respectively.
%

\section{Observations}
\label{s:obs}
Observations used to model the SED of V339~Del were collected 
from previous publications and satellite archives. Their timing, 
spectral range and sources are introduced in Table~\ref{tab:obs}. 

The X-ray fluxes of the super-soft source continuum were 
estimated using the RGS spectrum made with \textsl{XMM-Newton} 
at day 100. The spectrum was described by \cite{ness+13}. 
Because of a rich absorption spectrum and a strong 
C\,{\small V}$\rightarrow$C\,{\small VI} absorption edge 
at 31.6\,\AA, we adopted just the highest peaks in the 
observed spectrum as the continuum fluxes. We used the 
figure `P0728200201RGX000FLUXED1003.GIF' available in 
the \textsl{XMM-Newton} Science Archive 
(ID: 0728200201). 

The ultraviolet \textsl{HST/STIS} spectra were retrieved 
from the satellite archive with the aid of the Multimission 
Archive at the Space Telescope Science Institute (MAST). 
The spectra from days 35 and 100 were first reported by 
\cite{shore+13a,shore+13b} and published with a detailed 
description by \cite{shore+16} together with the spectrum 
from day 248. 

Low resolution optical spectra were carried out at the 
Crimean Astrophysical Observatory \citep[][]{tar+sk16}, 
at the Fujii Kurosaki Observatory (FKO) by M. Fujii\footnote{
http://otobs.org/FBO/fko/nova/nova$\_ $del$\_ $2013.htm} 
and from the \textsl{ARAS} database\footnote{
http://www.astrosurf.com/aras/Aras$\_ $DataBase/Novae/Nova-Del-2013.htm}. 
Spectroscopic observations from the FKO and \textsl{ARAS} were 
described by \cite{sk+14,sk+17}. 
Additional flux-points from the near-IR spectrum obtained at 
the Mt~Abu observatory were reconstructed from Fig.~7 of 
\cite{evans+17}. 
Spectroscopic observations were supplemented with the photometric 
$UBVJHKLM$ measurements of \cite{burlak+15}, $KLMNN'$ photometry of 
\cite{gehrz+15} and $BVR_{\rm C}I_{\rm C}$ magnitudes of 
\cite{munari+13b} and/or The \textsl{AAVSO} International 
Database\footnote{https://www.aavso.org/data-download}. 

To obtain flux-points of the true continuum using the $UBVR_C$ 
magnitudes, we corrected them for emission lines measured on 
the low-resolution spectra using the method of \cite{sk07}. 
The contribution of emission lines to the continuum was rather 
significant, mainly during the nebular phase, because of strong 
[O\,{\small III}] 4363\,\AA, 4950\,\AA, 5007\,\AA\ nebular lines 
and relative faint continuum (see Table~\ref{tab:dml}). 
Magnitudes of the true continuum were converted to fluxes 
according to the calibration of \cite{hk82} and \cite{bessell79}. 
Finally, the $UBVR_C$ continuum flux-points were used to scale 
the relative flux units of the low-resolution spectra to absolute 
fluxes. 
Observations were dereddened with $E_{\rm B-V}$ = 0.18 using 
the extinction curve of \cite{c+89}, and resulting parameters
were scaled to a distance of 4.5\,kpc (see Sect.~\ref{s:intro}). 
%
%
\begin{table}
\caption[]{Corrections $\Delta m_l = m_{\rm obs} - m_{\rm cont}$ 
           of the $UBVR_C$ photometry for emission lines, where 
           $m_{\rm cont}$ is magnitude of the true continuum. }
\begin{center}
\begin{tabular}{rcccc}
\hline
\hline
Day  & $\Delta U_l$$^a$ & $\Delta B_l$ & $\Delta V_l$ & $\Delta Rc_l$ \\
\hline
 35  & $<$-0.1  & -0.59  & -0.43  &  -1.18  \\
 45  & $<$-0.06 & -0.68  & -0.49  &  -1.00  \\
 59  & $<$-0.1  & -0.87  & -0.57  &  -1.27  \\
 68  & $<$-0.1  & -1.16  & -0.89  &  -1.36  \\
 72  & $<$-0.06 & -1.06  & -0.77  &  -1.32  \\
 77  & $<$-0.07 & -1.00  & -0.78  &  -1.28  \\
100  & $<$-0.13 & -0.87  & -0.68  &  -1.12  \\
113  & $<$-0.07 & -1.09  & -0.91  &  -1.32  \\
248  &  --      & -1.34  & -1.39  &  -1.11  \\
428  &  --      & -1.42  & -1.64  &  -1.18  \\
636  &  --      & -1.37  & -1.68  &  -1.64  \\
\hline
\end{tabular}
\end{center}
$^a$ Spectrum covers only a part of the $U$-passband range. 
\label{tab:dml}
\end{table}

\section{Analysis and results}
\label{s:analysis}
The primary aim of this paper is to model the SED of V339~Del 
throughout its iron-curtain phase (day 35), transition to 
the SSS phase (days 45, 59, 68, 72 and 77), the SSS phase 
(day 100, 113 and 248) and nebular phase (day 428 and 636). 
%
We achieve this aim by using the method of multiwavelength 
modeling the X-ray to IR composite continuum as described by 
\cite{sk05,sk15a}. Its application to the case of classical 
nova V339~Del is briefly introduced below. 
%
%
\begin{figure}
\begin{center}
\resizebox{\hsize}{!}{\includegraphics[angle=-90]{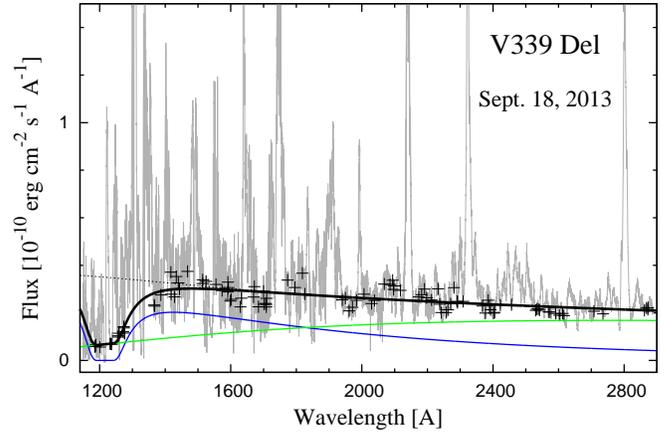}}
\end{center}
\caption{The \textsl{HST/STIS} spectrum of V339~Del on day 35 
showing an attenuation of the continuum around the Ly-$\alpha$ 
line due to Rayleigh scattering by atomic hydrogen 
(Sect.~\ref{sss:sed35}). 
The figure represents the UV part of the global SED from 
Fig.~\ref{fig:sed35}. Dotted line is the non-scattered 
continuum. Crosses denote the selected continuum fluxes 
used to model the SED. 
}
\label{fig:ray}
\end{figure} 

\subsection{Multiwavelength modeling the SED}
\label{ss:sed}
In our modeling we assume that the gamma photons generated in 
the burning layer on the WD surface are reprocessed by the 
outside material into softer energies, distribution of which 
throughout the electromagnetic spectrum depends on the nova age. 
Other primary sources of the radiation are not considered. 

Basic characteristics of the nova V339~Del, as inferred from its 
spectrum in different domains and dates (see Sect.~\ref{s:intro}), 
suggest that the continuum consists of three main components of 
radiation. The stellar component, $F_{\rm WD}(\lambda)$, produced 
by the WD pseudophotosphere, the nebular, $F_{\rm N}(\lambda)$, 
and dust, $F_{\rm D}(\lambda)$, component that represent a fraction 
of the stellar radiation reprocessed by the ejected material. 
The spectrum of the nova, $F(\lambda)$, as observed at the Earth, 
is given by their superposition, i.e., 
%
\begin{equation}
  F(\lambda) =  F_{\rm WD}(\lambda) + F_{\rm N}(\lambda) +
                F_{\rm D}(\lambda) .
\label{eq:sed1}
\end{equation}

During the fireball stage, the outer shell transfers the inner 
energetic photons to its optically thick/thin interface 
(i.e., the WD pseudophotosphere), which redistributes their 
energy chiefly within the optical. The observed spectrum can be 
compared with an atmospheric model for a star of spectral type 
A to F, while $F_{\rm N}(\lambda)$ and $F_{\rm D}(\lambda)$ 
components can be neglected \citep[see][]{sk+14}. 

During the transition to harder spectrum the 
WD pseudophotosphere shrinks and becomes hotter. 
As a result, the spectrum shifts the maximum of its SED to 
shorter wavelengths and the WD pseudophotosphere ionizes the 
outer material, giving rise to the nebular emission. 
Its continuum is approximated by contributions from f--b and 
f--f transitions in hydrogen plasma, while the stellar continuum 
and that from the dust are compared with the blackbody radiation 
at a temperature $T_{\rm BB}$ and $T_{\rm D}$, respectively. 

In the SED-fitting analysis, we compare a grid of models 
(\ref{eq:sed1}) for the given day (see below) with the fluxes 
of the observed continuum and select that corresponding to 
a minimum of the reduced $\chi^2$ function. Because of a rich 
emission line spectrum in the UV, it was difficult to identify 
its true continuum (see e.g., Fig.~\ref{fig:ray}). 
Therefore, we adopted uncertainties in the measured continuum 
as high as 10\%. 

\subsubsection{Modeling the iron-curtain spectrum on day 35}
\label{sss:sed35}
Observations used to model the SED of V339~Del at day 35 cover 
the spectral range from the far-UV to the near-IR 
(115\,nm to 5\,$\mu$m, Table~\ref{tab:obs}). Specific features 
of the observed spectrum are the flat UV continuum and its 
attenuation around the Ly-$\alpha$ line (Fig.~\ref{fig:ray}). 
The latter can be ascribed to the Rayleigh scattering of the 
far-UV photons by neutral atoms of hydrogen 
\citep[e.g.,][]{nussb+89}. It causes the optical depth, 
$\tau_{\rm Ray} = \sigma_{\rm Ray}(\lambda) N_{\rm H}$, where 
$\sigma_{\rm Ray}(\lambda)$ is the Rayleigh cross-section 
for scattering by hydrogen in its ground state 
\citep[see Eq.~(5) and Fig.~2 of][]{nussb+89}. 
Therefore, the attenuation of the continuum by Rayleigh 
scattering provides an estimate of $N_{\rm H}$ between 
the emitting source and the observer. 
According to the above-mentioned assumptions and Eqs.~(5) and 
(11) of \cite{sk05}, our Eq.~(\ref{eq:sed1}) can be written 
in the form, 
%
%
\begin{eqnarray}
 F(\lambda) =   
   \theta_{\rm WD}^2 \pi B_{\lambda}(T_{\rm BB})\,
    e^{-\sigma_{\rm Ray}(\lambda)\,N_{\rm H}} 
\nonumber \\
   +\, k_{\rm N} \times \varepsilon_{\lambda}({\rm H},T_{\rm e})
\nonumber \\
   +\, \theta_{\rm D}^2 \pi B_{\lambda}(T_{\rm D}),
\label{eq:sed2}
\end{eqnarray}
where 
$\theta_{\rm WD} = R_{\rm WD}^{\rm eff}/d$ is the angular radius 
of the WD pseudophotosphere, given by its effective radius 
(i.e. the radius of a sphere with the same luminosity) and the 
distance $d$. 
The second term at the right is the nebular continuum expressed 
by its volume emission coefficient, 
$\varepsilon_{\lambda}({\rm H},T_{\rm e})$ 
(${\rm erg\,cm^3\,s^{-1}\,\AA^{-1}}$), scaled with the factor 
$k_{\rm N}= EM/4\pi d^2$. The emission measure, 
$EM = \int_V n_{\rm p} n_{\rm e} {\rm d}V$, 
is given by the proton and electron concentration, $n_{\rm p}$ 
and $n_{\rm e}$ within the volume $V$ of the ionized hydrogen. 
Finally, the third term represents the radiation from the dust 
diluted at the Earth with the factor 
$\theta_{\rm D}^2 = (R_{\rm D}^{\rm eff}/d)^2$. 

\begin{figure*}[p!t]
\begin{center}
%
\resizebox{15cm}{!}{\includegraphics[angle=-90]{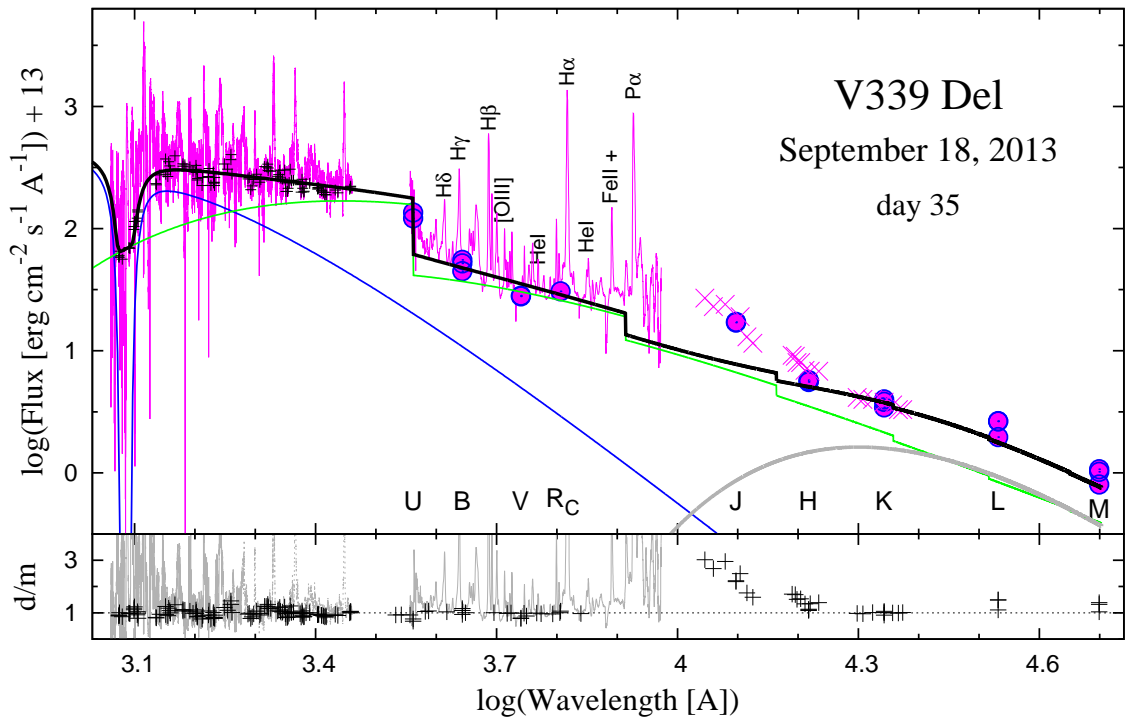}}
\end{center}
\caption{The observed (in magenta: spectrum and photometric 
flux-points) and model (heavy black line) SEDs of V339~Del 
from 0.12 to 5\,$\mu$m on day 35 with corresponding 
data-to-model ratios (d/m).
The model SED is given by a superposition of the radiation from
the WD pseudophotosphere (blue line), the nebula (green line) 
and the dust (gray line) according to Eq.~(\ref{eq:sed2}). 
The model unambiguously extracts an unidentified source of 
radiation around 1\,$\mu$m. 
}
\label{fig:sed35}
\end{figure*}

In modeling the global SED, we first fitted 105 UV continuum 
fluxes from 1185 to 2873\,\AA\ and 17 optical fluxes from 3412 
to 6918\,\AA\ by Eq.~(\ref{eq:sed2}) without the dust component. 
Second, we filled in the residual flux at $HKLM$ flux-points  
with a blackbody radiation to estimate parameters of the 
dust emission. 
In this way we obtained parameters $\theta_{\rm WD}$, 
$T_{\rm BB}$ and $N_{\rm H}$ for the stellar component, 
$T_{\rm e}$ and $k_{\rm N}$ for the nebular component and 
$\theta_{\rm D}$ and $T_{\rm D}$ for the dust component 
of radiation. 
The luminosity of the WD pseudophotosphere is given as 
%
\begin{equation}
  L_{\rm WD} = 4\pi d^2 \theta_{\rm WD}^2 \sigma T_{\rm BB}^4 .
\label{eq:lwd}
\end{equation} 
Finally, the model SED clearly extracts the unidentified 
source of emission around 1\,$\mu$m (see Fig.~\ref{fig:sed35}), 
whose presence was already noted by \cite{gehrz+15}. 
%
%

\subsubsection{Modeling the optical/near-IR spectrum during 
               the transition to the SSS phase}
\label{sss:sedust}
This period in the nova evolution is characterized with a steep 
decline of the optical brightness between day $\sim$35 and 
$\sim$72, and the presence of a strong near-IR emission. Our 
observations cover the nova age between day 45 and 113, and 
consist of low-resolution optical spectra and 
$UBVR_{\rm C}I_{\rm C}JHKLM$ photometry (Table~\ref{tab:obs}). 
During this period, the optical/near-IR continuum is dominated 
by the nebular emission and that from the dust. Therefore, 
we used these components to determine the model SED, 
\begin{equation}
 F(\lambda) = 
   k_{\rm N} \times \varepsilon_{\lambda}({\rm H},T_{\rm e}) 
   + \theta_{\rm D}^2 \pi B_{\lambda}(T_{\rm D}),
\label{eq:sedust}
\end{equation}
where the variables, $k_{\rm N}$, $T_{\rm e}$, $\theta_{\rm D}$ 
and $T_{\rm D}$ are explained in Sect.~\ref{sss:sed35}. 
Dominance of the nebular continuum and the dust emission in 
the optical and in the near-IR, respectively, allows us 
to determine both the components independently. 
Similarly to Eq.~(\ref{eq:lwd}) we determine the effective 
dust luminosity as 
$L_{\rm D} = 4\pi d^2 \theta_{\rm D}^2 \sigma T_{\rm D}^4$. 
Finally, we formally included the stellar component with 
$T_{\rm BB}\equiv 10^5$\,K and scaling, subtly influencing 
the short-wavelength part of the optical. 
Models are shown in Fig.~\ref{fig:sedir}. 
%
%
\begin{figure*}[p!t]
\begin{center}
\resizebox{15cm}{!}{\includegraphics[angle=-90]{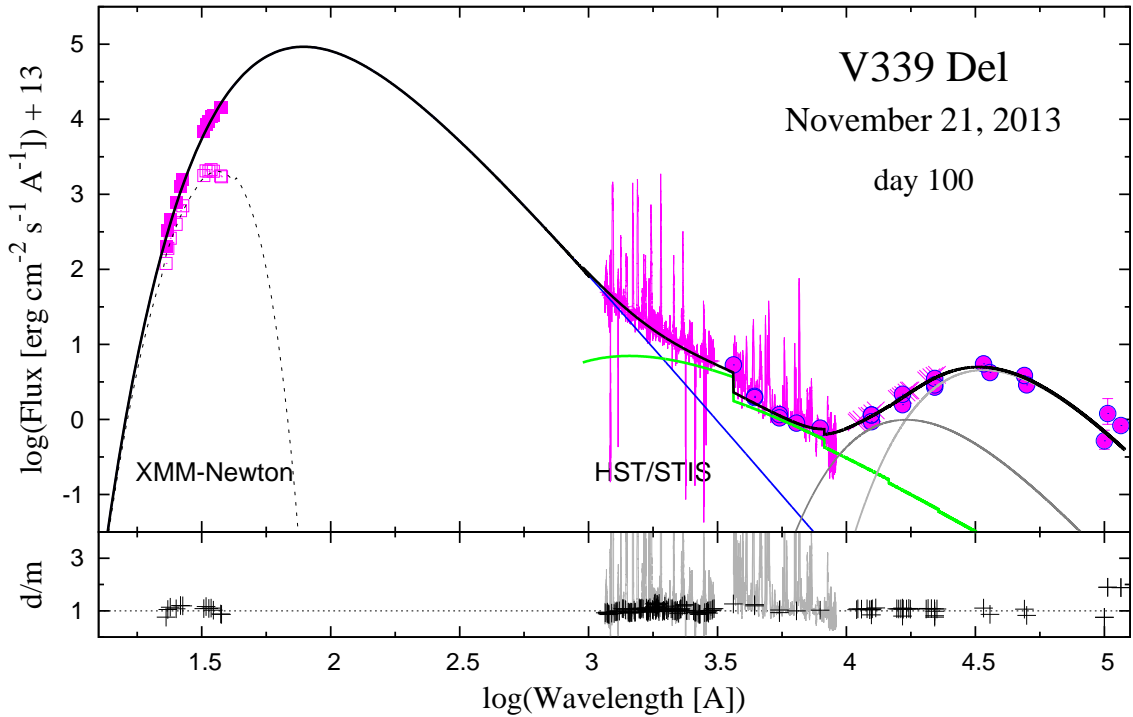}}
\end{center}
\caption{The observed and model SED of V339~Del from 2.3\,nm 
to 11\,$\mu$m on day 100. Open and filled squares are the 
measured and unabsorbed X-ray fluxes. Denotation of lines 
and points as in Fig.~\ref{fig:sed35}. 
}
\label{fig:sed100}
\end{figure*}

\subsubsection{Modeling the SSS spectrum on day 100 and 248}
\label{sss:sed100}
At the nova age of 100 days, we modeled its SED from the 
super-soft X-rays to the IR/N-band (2.3\,nm to 11\,$\mu$m, 
Table~\ref{tab:obs}). 
The spectrum is characterized with very high X-ray fluxes, 
a steep UV/optical continuum and a large emission bump 
in the IR. 
In principle, the procedure of modeling is the same as in 
the previous case, but the observed X-ray fluxes have to be 
corrected for bound--free absorptions within the interstellar 
(ISM) and circumstellar (CSM) matter. In this case the first 
term of Eq.~(\ref{eq:sed1}) is expressed as, 
%
\begin{equation}
  F_{\rm WD}(\lambda) =   
     \theta_{\rm WD}^2 \pi B_{\lambda}(T_{\rm BB})\,
     e^{-\sigma_{\rm X}(\lambda)\,N_{\rm H}}, 
\label{eq:sed3}
\end{equation}
where $\sigma_{\rm X}(\lambda)$ is the total cross-section 
for photoelectric absorption per hydrogen atom 
\citep[e.g.][]{crudd+74}. Here, we used the {\em tbabs} 
absorption model for ISM composition with abundances given 
by \cite{wilms+00} (e.g., $\log(A_{\rm CI})+12 = 8.38$). 

In the spectrum at day 100, we simultaneously fitted 13 absorbed 
X-ray continuum fluxes from 23 to 38\,\AA, 105 dereddened 
UV fluxes from 1161 to 3066\,\AA\ and 13 optical fluxes 
from 3554 to 8095\,\AA, searching for parameters 
$\theta_{\rm WD}$, $T_{\rm BB}$, $N_{\rm H}$, $T_{\rm e}$ 
and $k_{\rm N}$. 
The dust component was determined independently, because of 
a significant excess of its emission with respect to the nebular 
and stellar component in the IR. Comparing blackbody 
radiation to fluxes from 1.1\,$\mu$m to 11.6\,$\mu$m revealed 
the presence of two components of the dust emission. The larger 
and cooler component ($\lambda_{\rm max}\sim$3.4$\,\mu$m) was 
already identified by \cite{gehrz+15}, while the smaller and 
hotter one ($\lambda_{\rm max}\sim$1.7$\,\mu$m) is constrained 
by our global SED (see Fig.~\ref{fig:sed100}). 

During the late SSS phase on day 248, the {\em Swift}-XRT count 
rate was reduced by a factor of $\sim$1000 relative to day 100. 
The continuum used to model the SED was determined by the UV 
\textsl{HST/STIS} and optical spectra (1150 to 8200\,\AA, 
Table~\ref{tab:obs}). 
Strong far-UV continuum and its steep slope towards the longer 
wavelengths suggests a dominant contribution from a hot WD 
in the far-UV, while a relative flat and strong optical 
continuum constrains the presence of a hot nebular radiation. 
Therefore, we fitted the observed SED using Eq.~(\ref{eq:sed2}) 
restricted to the stellar and nebular component of radiation 
only, i.e., 
%
%
\begin{equation}
 F(\lambda) =   
   \theta_{\rm WD}^2 \pi B_{\lambda}(T_{\rm BB})\,
   +\, k_{\rm N} \times \varepsilon_{\lambda}({\rm H},T_{\rm e}), 
\label{eq:sed248}
\end{equation}
where variables $\theta_{\rm WD}$, $T_{\rm BB}$, $T_{\rm e}$ 
and $k_{\rm N}$ are described in Sect.~\ref{sss:sed35}. 
Here, we fitted 81 UV/optical continuum fluxes from 1159 
to 5620\,\AA\ of the \textsl{HST/STIS} spectrum with 
the function (\ref{eq:sed248}). In this way we determined 
parameters $\theta_{\rm WD}$, $T_{\rm e}$ and $k_{\rm N}$ 
for the fixed $T_{\rm BB}$. 
Under the condition that the radiation from the WD 
pseudophotosphere fits the UV/optical SED and generates 
the {\em Swift}-XRT count rate, it was possible to estimate 
the value of $T_{\rm BB}$ independently (see Appendix A in detail). 
For the observed rate of $\sim$0.025 photons s$^{-1}$ 
\citep[see Fig.~1 of][]{shore+16} and the flux, e.g., 
$F_{\rm WD}^{248}$(1195\,\AA) = 1.37$\times 10^{-12}$\ecsa, 
we obtained $T_{\rm BB}^{248} = 225\pm 5$\,kK. 
The model SED for day 248 is shown in Fig.~\ref{fig:seduvop}. 

\subsubsection{Modeling the nebular spectrum on day 428 and 636}
\label{sss:sed248}
During the nebular phase, the spectrum was modeled in the same 
way as that from day 248, because their SED profile was of 
the same type (see Fig.~\ref{fig:seduvop}). 
They differed only in the continuum level, which decreased 
with a factor of $\sim$3 and $\sim$7 on day 428 and 636 with 
respect to day 248. In both cases we estimated the lower limit 
of $T_{\rm BB}$ to $\sim 100\,000$\,K, as given by the presence 
of the N\,{\small V} 1238.8\,\AA, 1242.8\,\AA\ resonance doublet 
($\chi \sim 97.9$\,eV, see Sect.~\ref{ss:twd}), and its upper 
limit to $\sim 160\,000$\,K, at which the WD radiation is just 
outside of the {\em Swift}-XRT energy range. 

The observed and model SEDs are depicted in Figs.~\ref{fig:sed35}, 
\ref{fig:sed100}, \ref{fig:seduvop} and \ref{fig:sedir} and the 
corresponding parameters are listed in Table~\ref{tab:par}. 
In spite that only static blackbody-like models with nebular 
hydrogen continuum are used, the resulting models express 
well the measured {\em global} SED. 

\subsection{Temperature of the WD pseudophotosphere}
\label{ss:twd}
The WD temperature in novae depends on their age. 
Its determination is not often unique. Here, we introduce 
cases for day 35, 100 and 248, when the models SED allow 
to determine a reliable value of $T_{\rm BB}$. 

\subsubsection{$T_{\rm BB}$ on day 35}
\label{sss:twd35}
At the nova age of 35 days, the unusually flat UV continuum and 
the slope of the optical to near-IR continuum are given by 
superposition of the stellar radiation from a relatively 
warm WD pseudophotosphere and a large amount of a hot nebular 
radiation (Fig.~\ref{fig:sed35}, Table~\ref{tab:par}). 
Such the special profile of the UV--IR SED provides unambiguous 
solution. 
However, the observed WD radiation is not capable to give rise 
to the large amount of the nebular emission. 
The observed, $\sim$31\,000\,K warm, stellar pseudophotosphere 
(Table~\ref{tab:par}) generates the flux of hydrogen ionizing 
photons 
$L_{\rm H}\sim 9.5\times 10^{47}$\,s$^{-1}$, 
which can produce a maximum of 
$EM = L_{\rm H}/\alpha_{\rm B}({\rm H},T_{\rm e}) 
      \sim 8.1\times 10^{60}$\cmt\ 
(see Sect.~\ref{ss:lrwd} in detail) for the total hydrogen 
recombination coefficient 
$\alpha_{\rm B}({\rm H},T_{\rm e}) = 1.18\times 10^{-13}$ 
$\rm cm^{3}\,s^{-1}$ \citep[e.g.,][]{nv87}. 
However, this value is a factor of $\sim$23 below the measured 
quantity of $1.9\times 10^{62}$\cmt\ (Table~\ref{tab:par}). 
This implies that there is a powerful ionizing source in the center 
radiating at the ionization temperature $T^{i.s.} > T_{\rm BB}$, 
which is not seen directly by the outer observer at day 35. 

The lower limit of $T^{i.s.}$ is given by the presence of 
an ion in the nebula with the highest ionization energy 
$\chi_{\rm max}$, because it requires photons with 
$h\nu > \chi_{\rm max}$. According to \cite{mn94}, 
$T^{\rm i.s.} = 1000\times \chi_{\rm max}$, where $\chi_{\rm max}$ 
is in eV. The presence of 
N\,{\small IV} 1486\,\AA\ ($\chi \sim 77.5$\,eV) 
emission line 
suggests $T^{\rm i.s.} > 77\,000$\,K. 
On the other hand, absence of the 
N\,{\small V} 1238.8\,\AA, 1242.8\,\AA\ ($\chi \sim 97.9$\,eV)
resonance doublet in emission \citep[see][]{shore+16} 
suggests $T^{\rm i.s.} < 98\,000$\,K. 
Thus the temperature of the ionizing source on day 35 is in 
the range of 77\,000\,K -- 98\,000\,K. 
%
%
%
\begin{figure}[p!t]
\begin{center}
\resizebox{\hsize}{!}{\includegraphics[angle=-90]{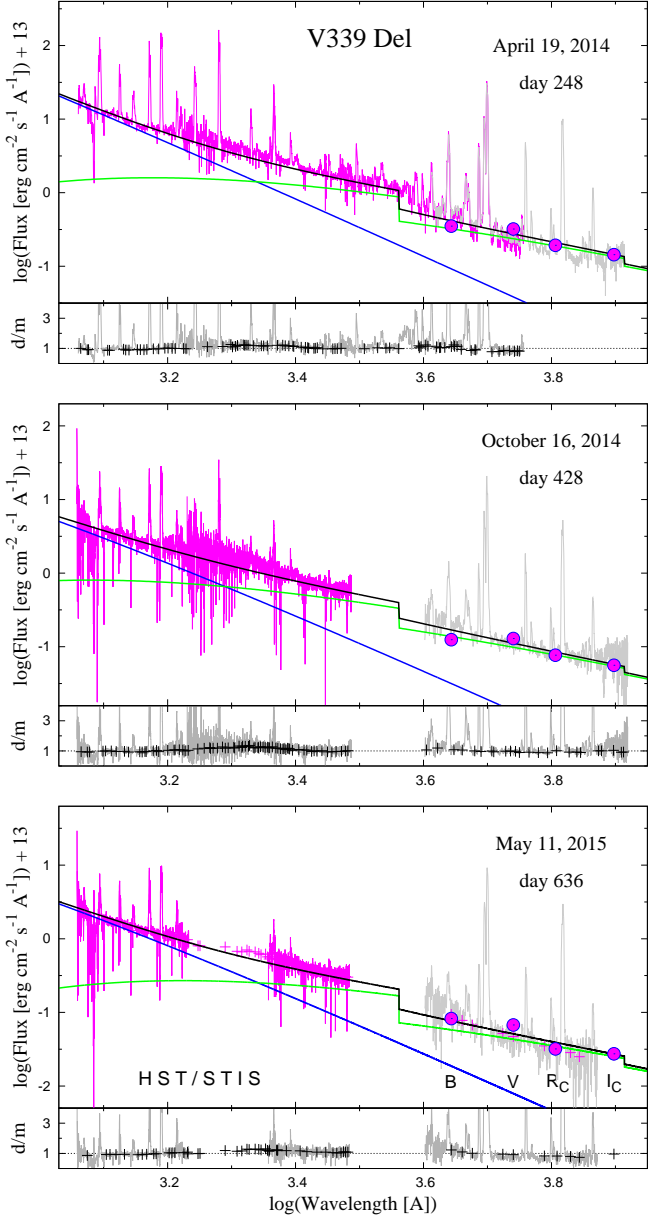}}
\end{center}
\caption{UV/optical SEDs on day 248, 428 and 636 of the nova 
age. Denotation of lines and points as in Fig.~\ref{fig:sed35}. 
}
\label{fig:seduvop}
\end{figure}
%
%
%
\begin{figure}
\begin{center}
\resizebox{\hsize}{!}{\includegraphics[angle=-90]{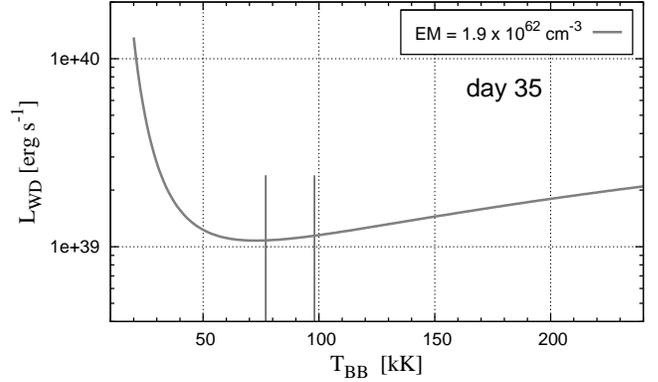}}
\end{center}
\caption{
WD luminosity as a function of the WD temperature 
scaled with the \textsl{EM} measured on day 35 
(Eq.~(\ref{eq:lwdem})). Vertical bars limit the range of 
the temperature (see Sect.~\ref{sss:twd35}). 
}
\label{fig:l35}
\end{figure}
\subsubsection{$T_{\rm BB}$ on day 100}
\label{sss:twd100}
At the nova age of 100 days, we determined $T_{\rm BB}$ directly 
by modeling the X-ray to IR continuum under the assumption that 
both X-ray and far-UV fluxes are emitted by 
the WD pseudophotosphere (see Sect.~\ref{sss:sed100}). 
Our value of $T_{\rm BB} = 369\pm 8$\,kK is close to the estimate 
of $\sim$350\,kK made by \cite{ness+13} on the basis of solely 
the X-ray fitting and assuming $N_{\rm H} = 1\times 10^{21}$\cmd. 
We note that determination of the effective temperature of the 
hot WD photosphere would require exact fitting of its rich 
absorption-line spectrum using a non-LTE model atmosphere 
analysis \citep[e.g.][]{lanz+05}. This challenging task is out 
of the scope of this paper. 
\subsubsection{$T_{\rm BB}$ on day 248 and beyond}
\label{sss:twd248}
%
During the late SSS phase (day 248), $T_{\rm BB}$ was determined 
by modeling the UV/optical SED with the aid of the measured 
{\em Swift}-XRT count rate (Sect.~\ref{sss:sed100}, 
Appendix~A). In this way, we obtained 
$T_{\rm BB} = 227\pm 5$\,kK. A relative small error is 
constrained by having defined the WD spectrum from both 
the short- and the long-wavelength side, similarly to day 100. 

During the nebular phase (day 428 and 636), only the range of 
possible $T_{\rm BB}$ can be estimated on the basis of 
the presence of emission lines with the highest ionization 
potential and the condition of no emission within the 
{\em Swift}-XRT energy range (Sect.~\ref{sss:sed248}). 
%
%
%
%
\begin{table*}[p!t]
\centering
\caption{Physical parameters of individual components of radiation 
         in the spectrum of V339~Del from models SED 
         (Sect.~\ref{ss:sed}). 
        }
\begin{tabular}{ccccccccccc}
\hline
                                              &
                                              &
\multicolumn{3}{c}{WD pseudophotosphere}      &
\multicolumn{2}{c}{Nebula}                    &
\multicolumn{3}{c}{Dust}                      & 
                                              \\
Age & &\multicolumn{8}{l}{-------------------------------------------~~~~~~
---------------------------~~~~~~~------------------------------------------}
&$\chi^2_{\rm red}$/d.o.f. \\
                                              & 
$N_{\rm H}$                                   & 
$T_{\rm BB}$                                  & 
$R_{\rm WD}^{\rm eff}$                        & 
$\log(L_{\rm WD})$                            & 
$T_{\rm e}$                                   & 
$EM$                                          & 
$T_{\rm D}$                                   &
$R_{\rm D}^{\rm eff}$                         & 
$L_{\rm D}$                                   & 
                                              \\
(d)                                           &
($10^{21}\,{\rm cm^{-2}}$)                    &
(kK)                                          &
($R_{\odot}$)                                 & 
(${\rm erg\,s^{-1}}$)                         & 
(kK)                                          & 
($10^{60}\,{\rm cm^{-3}}$)                    & 
(K)                                           & 
($R_{\odot}$)                                 & 
($L_{\odot}$)                                 & 
                                              \\
%
\hline\\
35   & 110$^{+20}_{-60}$ 
     &  31$\pm 1.5$
     & 5.9$\pm 0.8$
     &  38.05$\pm 0.11$ 
     & 26$\pm 2$ 
     & 190$\pm 15$ 
     & 1450$\pm 100$
     & 890$\pm 110$
     & 3120$\pm 700$
     & 1.69/121 \\
     & 
     & $77 - 98$
     & $1.7 - 1.1$
     & $39.0 - 39.1^{a}$
     & & & & & &  \\
45   &
     &
     &
     &
     & 30$\pm 5$ 
     & 84.8$\pm 9$ 
     & 1250$\pm 80$
     & 2760$\pm 250$
     & 16600$\pm 3200$
     & -- \\
59   &
     &
     &
     &
     & 30$\pm 5$ 
     & 14.5$\pm 2$ 
     & 1050$\pm 60$
     & 5200$\pm 500$
     & 29000$\pm 5500$
     & -- \\
68   &
     &
     &
     &
     & 40$\pm 7$ 
     & 9.7$\pm 1$ 
     & 1000$\pm 60$
     & 5030$\pm 480$
     & 22700$\pm 4400$
     & -- \\
72   & 
     & 
     & 
     & 
     & 50$\pm 10$
     & 8.48$\pm 0.9$ 
     & 900$\pm 70$
     & 5800$\pm 500$
     & 20100$\pm 3900$
     & -- \\
     & 
     & 
     &
     & 
     & 
     & 
     & 1800 $\pm 150$
     & 460 $\pm 50$
     & 2030 $\pm 350$
     &    \\
77   & 
     & 
     & 
     & 
     & 50$\pm 10$
     & 8.0$\pm 1$ 
     & 850$\pm 70$
     & 6200$\pm 530$
     & 17800$\pm 3400$
     & -- \\
     & 
     & 
     &
     & 
     & 
     & 
     & 1800 $\pm 150$
     & 530 $\pm 60$
     & 2630 $\pm 440$
     &    \\
100  & 1.02$\pm 0.1$
     & 369$\pm 8$
     & 0.20$\pm 0.02$
     & 39.43$\pm 0.11$
     & 50$\pm 10$
     & 8.48$\pm 0.8$
     & 850$\pm 70$
     & 5630$\pm 700$
     & 14840$\pm 3000$
     & 0.89/126 \\
     & 
     & 
     & 
     & 
     &
     & 
     & 1700$\pm 140$
     & 464$\pm 55$
     & 1600$\pm 330$
     &  \\
113  & 
     & 
     & 
     & 
     & 50$\pm 10$
     & 6.1$\pm 0.7$ 
     & 850$\pm 70$
     & 5300$\pm 450$
     & 13100$\pm 2600$
     & -- \\
     & 
     & 
     &
     & 
     & 
     & 
     & 1550 $\pm 130$
     & 560 $\pm 65$
     & 1600 $\pm 300$
     &    \\
248  & 0.8$\pm 0.4$
     & 227$\pm 5$
     & 0.16$\pm 0.01$
     & 38.36$\pm 0.12$
     & 48$\pm 8$
     & 1.96$\pm 0.5$
     & 
     & 
     & 
     & 1.50/77 \\
428  & 0.8$\pm 0.4$
     & 100$^{b}$
     & 0.15
     & 36.87
     & 60$\pm 10$
     & 0.91$\pm 0.1$
     &            
     &              
     &                
     & 1.27/94 \\
     & 
     & 160$^{b}$
     & 0.10  
     & 37.38
     & 60$\pm 10$   
     & 0.95$\pm 0.1$
     & 
     & 
     & 
     & 1.62/94 \\
636  & 0.8$\pm 0.2$
     & 100$^{b}$
     & 0.11
     & 36.64
     & 43$\pm 7$
     & 0.34$\pm 0.04$
     &            
     &              
     &                
     & 1.85/51 \\
     & 
     & 160$^{b}$
     & 0.08  
     & 37.12
     & 46$\pm 7$   
     & 0.39$\pm 0.04$
     & 
     & 
     & 
     & 2.27/51 \\
766  & 0.9$\pm 0.2$
     & 
     & 
     & 
     & 
     & 
     &           
     &           
     &           
     &  \\
867  & 0.8$\pm 0.3$
     &          
     &      
     &      
     &          
     &               
     &              
     &           
     &           
     &  \\ 
\hline
\multicolumn{10}{l}{{\bf Notes:}~
$^{a})$ from the measured \textsl{EM} 
        (see Sects.~\ref{ss:twd} and \ref{ss:lrwd}), 
$^{b})$ fixed value (see Sects.~\ref{ss:twd})} 
\end{tabular}
\label{tab:par}
\end{table*}

\subsection{Luminosity of the burning WD}
\label{ss:lrwd}      
The bolometric luminosity, $L_{\rm WD}$, of the hot burning WD 
at day 35 can be estimated only indirectly, under the assumption 
that the total flux of hydrogen ionizing photons, $L_{\rm H}$, 
is converted into the nebular radiation. Then having the quantity 
of \textsl{EM} from observations we can estimate the corresponding 
$L_{\rm WD}$ for the given temperature of the ionizing source. 

Assuming further that 
the nebula is characterized with a constant electron temperature, 
$T_{\rm e}$, we can approximate the equilibrium condition between 
$L_{\rm H}$ and the rate of recombinations in the nebula as 
\begin{equation}
   L_{\rm H} = \alpha_{\rm B}({\rm H},T_{\rm e})\,\textsl{EM},
\label{eq:lH}
\end{equation}
where $\alpha_{\rm B}({\rm H},T_{\rm e})$ is the recombination 
coefficient to all but the ground state of hydrogen 
(i.e., Case $B$). 
Quantity of $L_{\rm H}$ is given by the temperature and luminosity 
of the ionizing source, while the value of \textsl{EM} is
determined by the model SED. 
According to \cite{sk+17} the luminosity of the ionizing source, 
which produces the observed \textsl{EM}, can be expressed as 
\begin{equation}
  L_{\rm WD} = \alpha_{\rm B}({\rm H},T_{\rm e})\,\textsl{EM}
               \frac{\sigma T_{\rm BB}^{4}}{f(T_{\rm BB})},   
\label{eq:lwdem}
\end{equation}
where the function $f(T_{\rm BB})$ determines the flux of 
ionizing photons emitted by 1\,cm$^2$ area of the ionizing 
source (cm$^{-2}$\,s$^{-1}$). 
Figure~\ref{fig:l35} shows Eq.~(\ref{eq:lwdem}) 
for the observed \textsl{EM} = $1.9\times 10^{62}$\cmt. 
The temperature range of the ionizing source, 
77\,000\,K -- 98\,000\,K (Sect.~\ref{ss:twd}), corresponds 
to $L_{\rm WD} \sim 1.1\times 10^{39}$\es. 

Optically thick conditions, where the nebular continuum 
growths from complete absorption of stellar photons with 
$\lambda < 912$\,\AA, are of critical importance for the 
validity of the above approach. 
A signature of the optically thick conditions is the presence 
of elements at very different ionization states in the 
spectrum \citep[see][ and references therein]{kj89}. 
On day 35, emission lines of N\,{\small II--IV}, 
C\,{\small I--IV}, Si\,{\small IV}, He\,{\small I--II}, 
H\,{\small I}, O\,{\small I}, [O\,{\small I}]\,$\lambda 6300$ 
and Fe\,{\small II} as indicated in the UV--optical spectrum 
of V339~Del \citep[see Figs.~3, 8 and A.1. of][]{shore+16} 
suggest that the nebula is rather ionization-bounded, that is, 
the stellar radiation below 912\,\AA\ ($<$13.6\,eV) is absorbed 
within the nebula. Also, no detection of ionizing photons on 
day 35 within the {\em Swift}-XRT energy range 
\citep[see Fig.~1 of][]{shore+16} supports this case. 

During the SSS phase, direct observation of a significant 
fraction of the hydrogen ionizing photons in the form of 
super-soft X-ray emission (Fig.~\ref{fig:sed100}) means 
that the nebula of V339~Del became to be particle-bounded. 
As a result its \textsl{EM} significantly dropped with 
respect to the value from day 35 (Table~\ref{tab:par}). 
The method above (Eq.~(\ref{eq:lwdem})) is not applicable 
for the spectrum from day 100. 
Instead, we can directly integrate the stellar component of 
radiation given by the X-ray/near-IR model SED, which yields 
$L_{\rm WD} = (2.7\pm 0.7)\times 10^{39}(d/4.5\,{\rm kpc})^2$\es. 
This value justifies those obtained indirectly using 
the nebular component of radiation. 

Luminosities during the late SSS and nebular phase relay on 
estimates of the corresponding $T_{\rm BB}$ 
(see, Sect.~\ref{ss:twd}). 
Figure~\ref{fig:evoll} shows evolution of the WD luminosity 
as a function of the nova age. It demonstrates a long-lasting 
super-Eddington luminosity of the burning WD in V339~Del at least 
for the first 100 days of its life. 
%
%
\begin{figure}
\begin{center}
%
\resizebox{\hsize}{!}{\includegraphics[angle=-90]{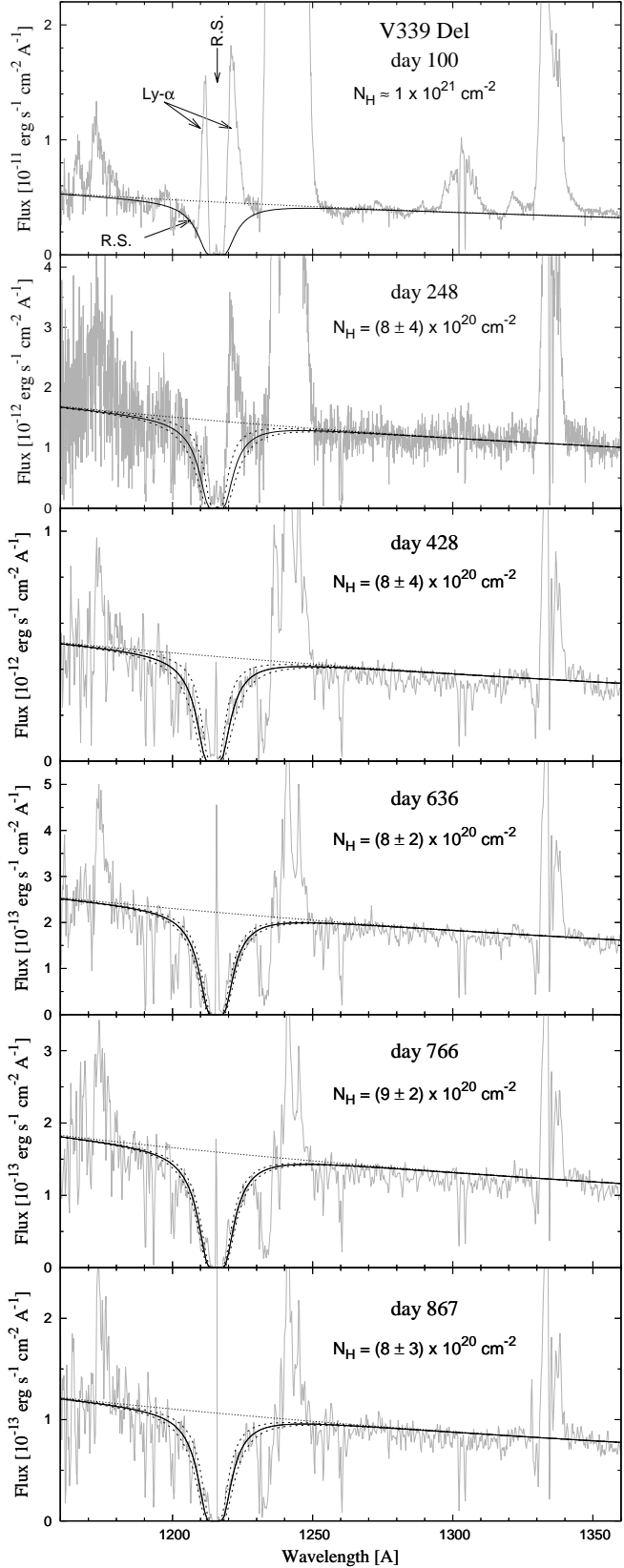}}
\end{center}
\caption{The hollow in the continuum around Ly-$\alpha$ observed 
during the SSS and nebular phase and its match by Rayleigh 
scattering (full and dashed lines, R.S.). Dotted lines 
above the hollow represent the non-scattered continuum. 
}
\label{fig:rayall}
\end{figure}

\subsection{Veiling of the WD by neutral hydrogen}
\label{ss:nh}
Apart from b--f absorptions of X-ray photons by hydrogen atoms, 
the hydrogen column density between the WD pseudophotosphere and 
the observer, $N_{\rm H}$, can be also determined from attenuation 
of the far-UV continuum around the Ly-$\alpha$ line, which is 
caused by Rayleigh scattering on H atoms (Sect.~\ref{sss:sed35}). 
On day 35, the continuum depression around Ly-$\alpha$ was 
stretched up to $\sim$1400\,\AA\ (Fig.~\ref{fig:ray}), which 
corresponds to 
$N_{\rm H} = 1.1^{+0.2}_{-0.6}\times 10^{23}$\cmd, in agreement 
with the following estimates from X-ray observations 
(Fig.~\ref{fig:nhem}). 
Such the high value of $N_{\rm H}$ is in major part given by 
its CSM component. 
It is of interest to note that signatures of Rayleigh scattering, 
corresponding to similar values of $N_{\rm H}$, developed also 
in the spectra of the symbiotic nova PU~Vul and classical nova 
V723~Cas during their transition to harder spectrum 
\citep[see Fig.~1 of][]{sk14}. 

On day 100, having observed the super-soft X-ray component 
of the nova spectrum, we determined $N_{\rm H}$ directly by 
modeling the SED (Sect.~\ref{sss:sed100}) to 
$\sim 1.0\times 10^{21}$\cmd. 
This value corresponds to the interstellar component, because 
it is relevant to the extinction to the nova, 
$E_{\rm B-V} = 0.18$\,mag \citep[e.g.,][]{d+s94}. 
Comparison of the far-UV spectrum with the Rayleigh scattered 
model continuum also supports this value of $N_{\rm H}$, although 
the influence of the Ly-$\alpha$ emission is significant 
(see the top panel of Fig.~\ref{fig:rayall}). 

During the following observations (day 248, 428, 636, 766 and 867) 
attenuation of the far-UV continuum by Rayleigh scattering was 
clearly recognizable, corresponding just to the ISM component 
of $N_{\rm H} \sim 8\times 10^{20}$\cmd\ with uncertainties up 
to 50\% (Table~\ref{tab:par}, Fig.~\ref{fig:nhem}), depending on 
the S/N ratio\footnote{To suppress the noise, we smoothed the spectra 
               using average values of fluxes within 0.25\,\AA.}. 
Therefore, in contrast to Eq.~(\ref{eq:sed2}), we attenuated 
the total model continuum, $F(\lambda)$, with the function 
exp$[-\sigma_{\rm Ray}(\lambda)\,N_{\rm H}]$, because the only 
present ISM component of $N_{\rm H}$ attenuates both the stellar 
and the nebular radiation from the nova. As a result, the rest 
flux around the reference wavelength is close to zero 
(see Sect.~\ref{ss:ray}). 

Our fits of the continuum attenuation around Ly-$\alpha$ by 
Rayleigh scattering are shown in Fig.~\ref{fig:rayall}, while 
the evolution of $N_{\rm H}$ along the nova age is depicted 
in Fig.~\ref{fig:nhem}. 

\section{Discussion}
\label{s:dis}
\subsection{Transition from the iron-curtain to the SSS phase}
\label{ss:transition}
Here we discuss the steep decline of the optical brightness, 
\textsl{EM} and $N_{\rm H}$ 
between day 35 and 72. With the aid of models SED we endeavor 
to find its origin. 

\subsubsection{Stopping the mass-outflow from the WD}
\label{sss:mdot}
Models SED show that the nebular component of radiation dominates 
the near-UV to optical spectrum for the nova age $\gtrsim$35 days 
(Figs.~\ref{fig:sed35}, \ref{fig:sed100}, \ref{fig:seduvop} and 
\ref{fig:sedir}). 
It represents a fraction of the WD's radiation converted by the 
ejecta into the nebular radiation via f--f and f--b transitions. 
Its flux is given by the \textsl{EM} of the ionized gas 
(see Sect.~\ref{sss:sed35}, Eq.~(\ref{eq:sed2})). 
Therefore, knowing the \textsl{EM} from the model SED, we can 
estimate the mass of the ionized ejecta and the corresponding 
mass-loss rate from the WD, $\dot M_{\rm WD}$. 
According to the simplified approach of \cite{sk+14}, 
$\dot M_{\rm WD}$ can be expressed as, 
%
\begin{equation}
 \dot M_{\rm WD} =
         \mu m_{\rm H} v_{\rm exp}\left[\epsilon\,4\pi 
         \textsl{EM}\left(\frac{1}{R_{\rm WD}^{\rm eff}} -
         \frac{1}{R_{\rm neb}}\right)^{-1} 
         \right]^{1/2}  {\rm g s^{-1}},    
\label{eq:mdot1}
\end{equation}  
where $\mu$ is the mean molecular weight and $m_{\rm H}$ 
the mass of the hydrogen atom, while $v_{\rm exp}$ denotes 
the expansion velocity of the ejecta, 
$\epsilon = 2\Delta\Omega/4\pi$ is the filling factor, where 
$\Delta\Omega < \pi$ is the opening angle of the expanding 
ionized region, whose \textsl{EM} is integrated from 
$R_{\rm WD}^{\rm eff}$ to the outer radius of the nebula,
$R_{\rm neb}$. 

\begin{figure}[p!t]
\begin{center}
\resizebox{\hsize}{!}{\includegraphics[angle=-90]{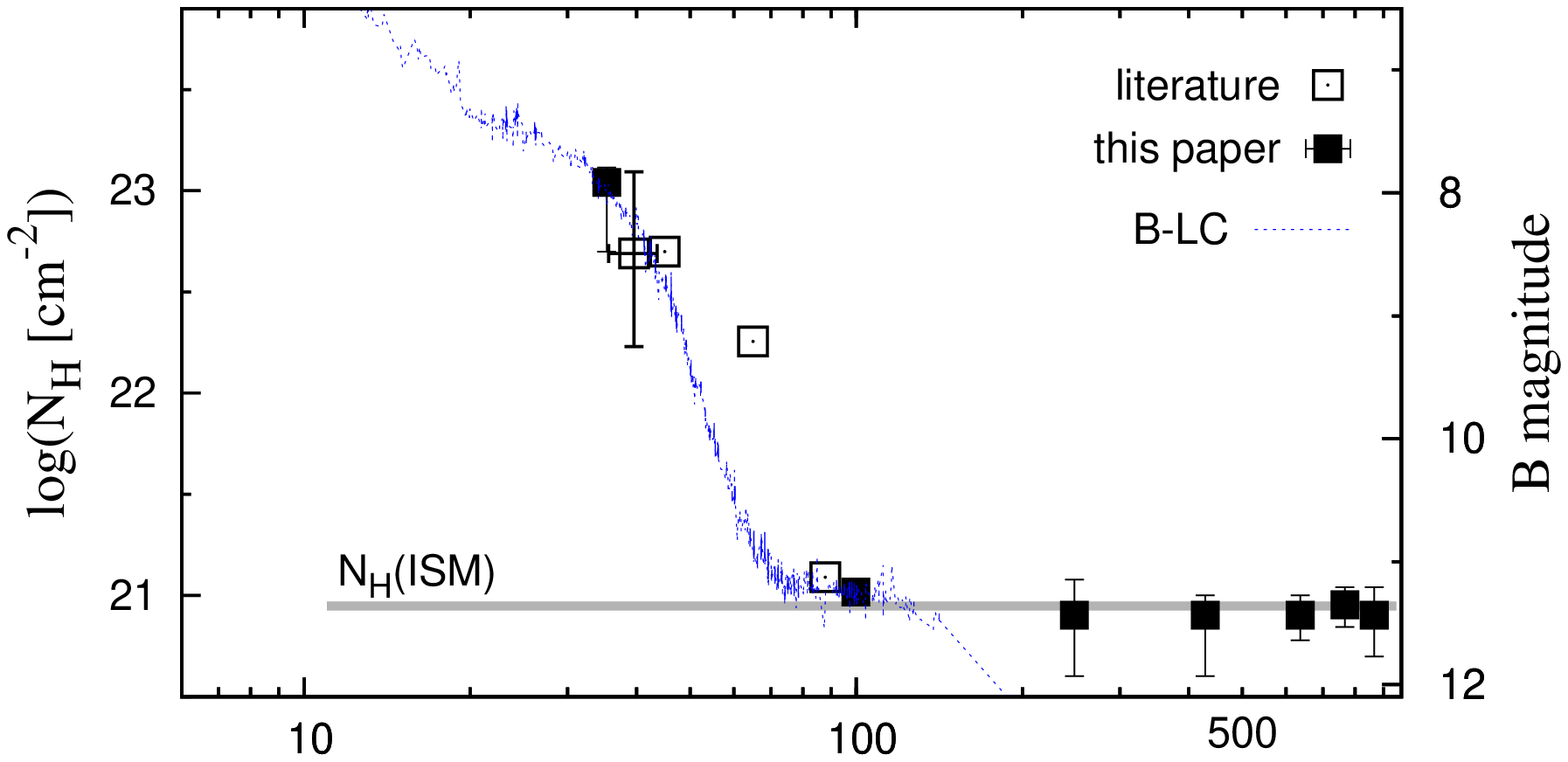}}\vspace*{-3mm}
\resizebox{\hsize}{!}{\includegraphics[angle=-90]{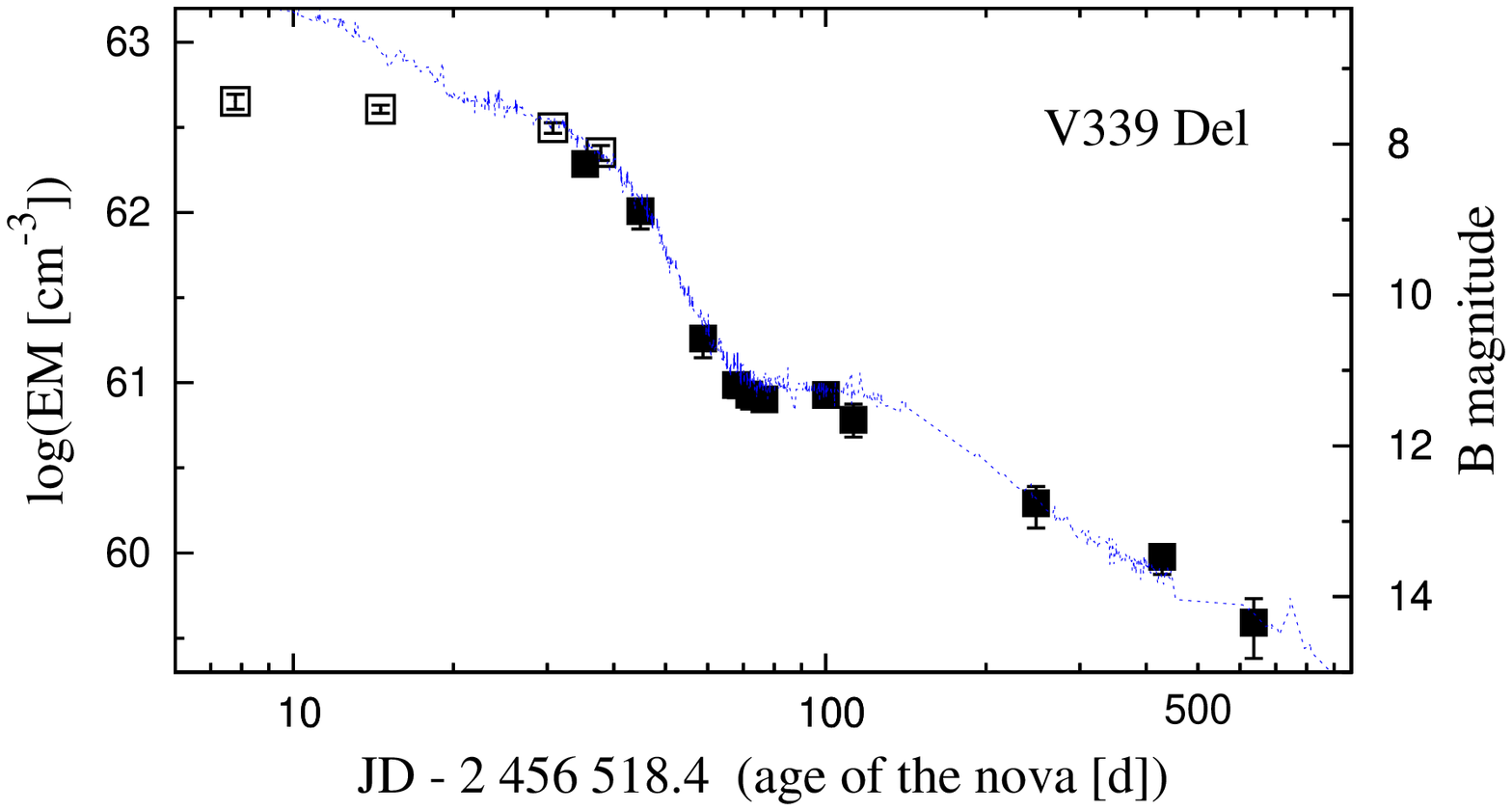}}
\end{center}
\caption{
Top: Column density of hydrogen atoms, $N_{\rm H}$, along 
the nova age. Horizontal line corresponds to the ISM value 
(Sect.~\ref{ss:nh}). 
Bottom: As in the top, but for the \textsl{EM}. 
The $B$-LC (blue line) is scaled to both parameters 
on y2-axis (Sect.~\ref{sss:drop}). 
Our data are from Table~\ref{tab:par}. 
}
\label{fig:nhem}
\end{figure}

For a mean expansion velocity of 750\,\kms\ derived from 
the \textsl{HWHM} of the hydrogen emission lines 
\citep[][]{tar+sk16,evans+17} and parameters of the model SED, 
$R_{\rm WD}^{\rm eff}$ and \textsl{EM} (Table~\ref{tab:par}), 
we obtain $\dot M_{\rm WD} = 8.7\times 10^{-5}$\myr\ for 
day 35, while on day 72, when the drop in the brightness and 
in the \textsl{EM} came to a stand-still (Fig.~\ref{fig:nhem}), 
$\dot M_{\rm WD}$ decreased by a factor of $\sim$25, to 
$3.4\times 10^{-6}$\myr. We assumed 
$R_{\rm WD}^{\rm eff} \ll R_{\rm neb}$ and $\epsilon = 1$, 
which corresponds to a maximum of $\dot M_{\rm WD}$. 

Thus, the rapid decline of the \textsl{EM} 
during the brightness fall ($\sim$days~35--72) was caused by 
a significant stopping-down the mass-outflow from the WD. 

\subsubsection{On the nature of the steep brightness decline}
\label{sss:drop}
This period in the nova evolution is characterized by 
a significant decrease of its optical brightness by 
$\Delta B\approx 3$\,mag and $\Delta I_{\rm C}\approx 4$\,mag 
between day 35 and day 72 (see Fig.~\ref{fig:lc}). 
Simultaneously, $N_{\rm H}$ decreased from $\sim 1\times 10^{23}$ 
to its ISM value of $\sim 1\times 10^{21}$\cmd\ and \textsl{EM} 
decreased from $\sim 2\times 10^{62}$ to 
$\sim 8.5\times 10^{60}$\cmt\ (Fig.~\ref{fig:nhem}). 

According to models SED (Figs.~\ref{fig:sed35}, \ref{fig:sed100} 
and \ref{fig:sedir}), the fall of the optical brightness prior 
to the SSS phase was caused by the decrease of both the 
nebular and the stellar continuum. A distinctive {\em increase} 
in $T_{\rm BB}$ and $T_{\rm e}$ to day 100 shifted the maximum 
of the stellar component to shorter wavelengths and made both 
the continua much steeper than on day 35, which yields 
$\Delta I_{\rm C} > \Delta B$. 
This is because the nebular component, 
$F_{\rm N}(\lambda) = EM/4\pi d^2 \times 
                      \varepsilon_{\lambda}({\rm H},T_{\rm e})$ 
(Sect.~\ref{sss:sed35}), dominates the optical, and thus we can 
estimate the value of the drop as, 
%
\begin{equation}
  \Delta m_{\lambda} = -2.5\log\left[\frac{EM^{35}}{EM^{72}}
        \frac{\varepsilon_{\lambda}^{35}({\rm H},T_{\rm e})}
             {\varepsilon_{\lambda}^{72}({\rm H},T_{\rm e})}
              \right],
\label{eq:dm}
\end{equation}
where indices 35 and 72 denote quantities corresponding to 
these days of the nova age. For values of \textsl{EM}, 
$T_{\rm e}$ in Table~\ref{tab:par} and the continuum-emission 
coefficient $\varepsilon_{\lambda}({\rm H},T_{\rm e})$ 
introduced by, e.g., \cite{brown70}, Eq.~(\ref{eq:dm}) yields 
$\Delta B = 3.5\pm 0.05$ and $\Delta I_{\rm C} = 3.8\pm 0.05$\,mag, 
in agreement with the observed values. 

\begin{figure}[p!t]
\begin{center}
\resizebox{\hsize}{!}{\includegraphics[angle=-90]{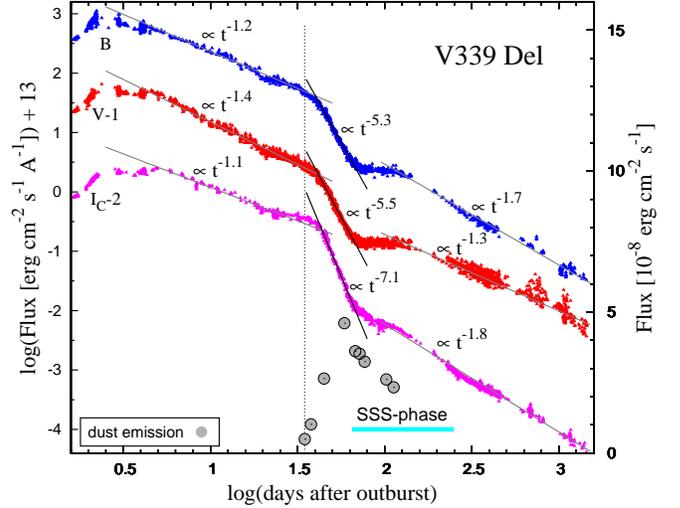}}
%
\end{center}
\caption{
Fluxes corresponding to $B,V,I_{\rm C}$ magnitudes according to 
conversion of \cite{hk82}. Power-law fits are for 5--40, 42--62 
and $\sim$150--1500 day parts of the LC. Measured flux of the 
dust emission is scaled on y2-axis (data from Table~\ref{tab:par}). 
Dotted line denotes its first detection. 
}
\label{fig:slopes}
\end{figure}

Simultaneous fading of both the $N_{\rm H}$ and \textsl{EM} between 
day 35 and 72 was caused by a drop in the $\dot M_{\rm WD}$
(Sect.~\ref{sss:mdot}), which considerably lowers particle 
concentration of the nebula, and thus its absorbing ability 
and emissivity. 
As a result, the ejecta became optically thinner for harder 
photons, which led to shrinking of the WD pseudophotosphere, 
the increase of its temperature and, consequently, an increase 
of the electron temperature of the irradiated nebula 
(Table~\ref{tab:par}). Therefore, the very steep optical 
brightness decline with the power of -5 to -7 in fluxes 
(see Fig.~\ref{fig:slopes}) was in part caused by shifting 
of both components of radiation from the optical to shorter 
wavelengths. The nova settled at the SSS phase. 
%
%
%

\subsubsection{Is there a dust extinction along the line of sight?}
\label{sss:dust}
\begin{figure}
\begin{center}
%
\resizebox{7.8cm}{!}{\includegraphics[angle=-90]{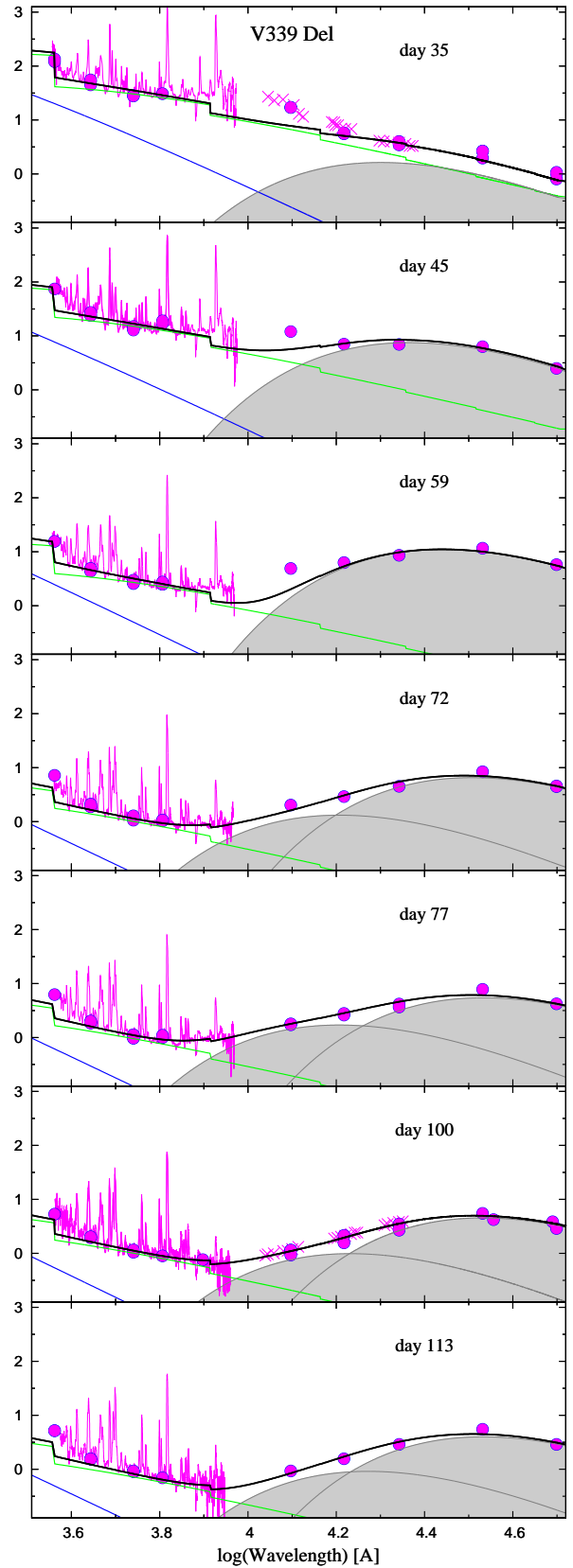}}
\end{center}
\caption{
Optical/near-IR SEDs during the transition from the iron-curtain 
to the SSS phase, when a strong dust emission developed 
(gray area; Sect.~\ref{sss:sedust}). 
Denotation of lines and points as in Fig.~\ref{fig:sed35}. 
Fluxes are in $10^{-13}$\ecsa. 
}
\label{fig:sedir}
\end{figure}

Based on models SED we suggested that the simultaneous drop 
in the stellar brightness, \textsl{EM} and $N_{\rm H}$, during 
the transition from day 35 to the SSS phase, is caused by 
stopping-down the mass-loss from the WD (Sects.~\ref{sss:mdot} 
and \ref{sss:drop}). 
On the other hand, \cite{evans+17} ascribed the decline in the 
visual LC to dust extinction. 
However, following facts do not support this view. 
%
\begin{enumerate}
\item
The observed fall of the flux in the $I_{\rm C}$ band 
($\lambda_{\rm eff} = 7900$\,\AA) is steeper and larger 
than in the $B$ band ($\lambda_{\rm eff} = 4400$\,\AA). 
Figure~\ref{fig:slopes} demonstrates this case. The slope 
of flux $F_{\rm Ic}\propto t^{-7.1}$, while 
$F_{\rm B}\sim F_{\rm V} \propto t^{-5.4}$, which corresponds to 
$\Delta I_{\rm C} > \Delta B (\Delta V)$ during the flux fall 
(see also Eq.~(\ref{eq:dm})). 
Such the behavior contradicts attenuation of the light by 
the dust \citep[e.g.,][]{c+89}. 
For a comparison, the wavelength dependence of dust extinction 
was clearly demonstrated for the nova V5668~Sgr 
\citep[see Fig.~3 of][]{gehrz+18}. 
\item
No dust extinction along the line of sight is indicated by 
multiwavelength modeling the SED before, during and after 
the steep flux decline. 
All the corresponding models (see Figs.~\ref{fig:sed35},
\ref{fig:sed100} and \ref{fig:sedir}) fit observations 
corrected solely with the ISM component of the dust extinction, 
$E_{\rm B-V} = 0.18$\,mag. 
\item
The IR-excess produced by the dust emission was clearly 
measured already on day 34.8, i.e., around 6--7 days prior 
to the steep flux decline (Fig.~\ref{fig:slopes}). 
\end{enumerate}
\cite{shore+18} suggested that a short-lasting dip in the 
{\em Swift}-UV LC around day 77 (see their Fig.~11) signals 
formation of dust, although there is no counterpart seen in 
the optical LC \citep[see e.g., precise LCs of][]{munari+15}. 
However, our optical/near-IR models SED indicate creation of 
the dust more than one month prior to the dip in the $UV$ 
continuum. A rapid growth of the near-IR emission happened 
already between day 35 and 45, with a maximum around day 59 
(Figs.~\ref{fig:slopes} and \ref{fig:sedir}). 
It is of interest to note that the $UV$ dip coincides with 
emergence of the rapid variation in the {\em Swift}-XRT count 
rate \citep[see Fig.~1 of][]{shore+16}, which suggests their 
common origin. According to models SED, the $X$-ray to $UV$ 
continuum ($\lambda\lesssim$2000\,\AA) is dominated by the 
radiation from the WD photosphere, and thus these events 
probably reflect its variability. However, the optical is 
contaminated by the nebular continuum, which overlays the 
presumable variation of the stellar component. 
This suggests that the agent (dense clumps?) responsible 
for the changes in the $X$-ray/$UV$ continuum is located 
above the ionizing source, but below the extended nebula 
(\textsl{EM}$\sim 10^{61}$\cmt), whose light cannot vary 
on the timescale of days. 

To answer the question of this section, it is also important 
to consider the effect of hard radiation on the survival of 
dust in the nova environment. 
This effect was discussed in detail by \cite{fruchter+01}, who 
considered grain heating and grain charging as two principal 
mechanisms responsible for dust destruction. 
Using their approach, \cite{evans+17} found that charging of 
grains by $X$-radiation is more than sufficient to shatter 
the grains around V339~Del already before day 100. 
Also \cite{gehrz+18} showed that the $X$-ray fluence is 
sufficient to destroy the dust in the nova V5668~Sgr if 
the grains are exposed to X-rays for $\sim$1 month. 
On the other hand, to explain the asymmetry in the line profiles 
with a suppressed red wing, \cite{shore+18} state that there 
is no need to invoke dust destruction at late times in either 
V5668~Sgr or V339~Del, even under irradiation by the $X$-ray 
and EUV from the central star\footnote{However, a depression 
of the red wings due to blocking a fraction of the radiation 
from the receding ejecta by the disk 
\citep[see Sect.~4.1. of][]{sk+06} was not considered.}. 

According to the arguments above (points 1. to 3. of this section) 
and our interpretation (Sect.~\ref{sss:drop}), we conclude that 
there is no detectable dust extinction along the line of sight 
during the steep drop in the star's brightness. 
Hence, the presence of a strong near-IR emission throughout this 
period (Fig.~\ref{fig:sedir}) and far beyond it \citep[][]{evans+17} 
requires a non-spherical arrangement of the dust region within 
the nova ejecta. 
%
%
%

\subsection{On the structure of the nova ejecta}
\label{ss:structure}
Already two days after the explosion, observations indicated 
a prolate structure of the nova. Later evolution suggested 
a biconical, disk-polar ionization structure of the ejecta 
(see Sect.~\ref{s:intro}). 
Here, Fig.~\ref{fig:sketch} shows a sketch for the nova 
ejecta as can be inferred from models SED on day 35 and 100, 
just before and after the sudden drop in the star's brightness. 
It is similar to that derived from radio imaging of V959~Mon 
\citep[see Fig.~3 of][]{chomiuk+14}, specifying some details 
of its inner part. 

\subsubsection{The oblate shape of the WD pseudophotosphere}
\label{sss:wd}
On day 35, the flat UV continuum is a result of superposition 
of a relatively warm stellar and a strong nebular 
component of radiation (Fig.~\ref{fig:sed35}). 
The former is not capable of giving rise to the observed nebular 
emission and thus the latter signals the presence of a strong 
ionizing source in the system with $T^{\rm i.s.} > T_{\rm BB}$, 
which is not seen directly by the observer 
(see Sect.~\ref{sss:twd35}). 
Such the type of the spectrum is called the two-temperature-type 
UV spectrum. It is often observed during Z~And-type outbursts 
of symbiotic binaries, where it is followed by emergence of 
an attenuation of the far-UV continuum by Rayleigh scattering 
on a few times $10^{22}-10^{23}$ hydrogen atoms 
\citep[see Sect.~5.3.4., Fig.~27 and Table~4 of][]{sk05}. 
The corresponding SED is explained by a disk-like structure of 
the WD pseudophotosphere, whose hotter regions with smaller radii 
are located at/around its poles and vice versa, cooler regions 
with larger radii are located towards the WD equator. 
 
Accordingly, the day 35 model SED of V339~Del suggests such 
the oblate shape of the WD pseudophotosphere 
(see Fig.~\ref{fig:sketch}, top) with relevant non-spherical 
temperature distribution: the observed $\sim$31\,kK warm 
pseudophotosphere and 77 -- 98\,kK hot central ionizing source 
(Sect.~\ref{sss:twd35}). 
A higher density stellar wind in the direction of the observer 
creates the optically thick/thin interface (i.e., the WD 
pseudophotosphere) at a larger distance from the wind origin 
than at its pole. The wind gives rise to a large amount 
of CSM above it as indicated by the high value of $N_{\rm H}$, 
which precludes detection of the super-soft X-rays on day 35. 

\subsubsection{Dusty disk during the SSS phase}
\label{sss:ddisk}
\begin{figure}[p!t]
\begin{center}
\resizebox{\hsize}{!}{\includegraphics[angle=-90]{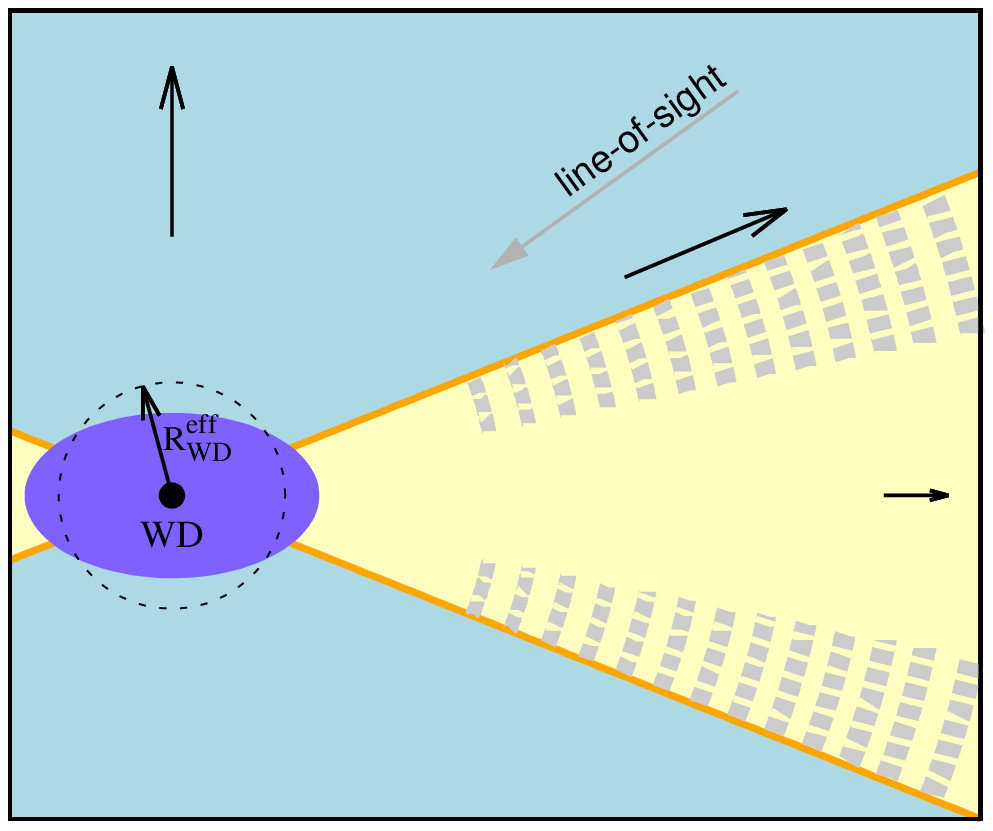}}
\resizebox{\hsize}{!}{\includegraphics[angle=-90]{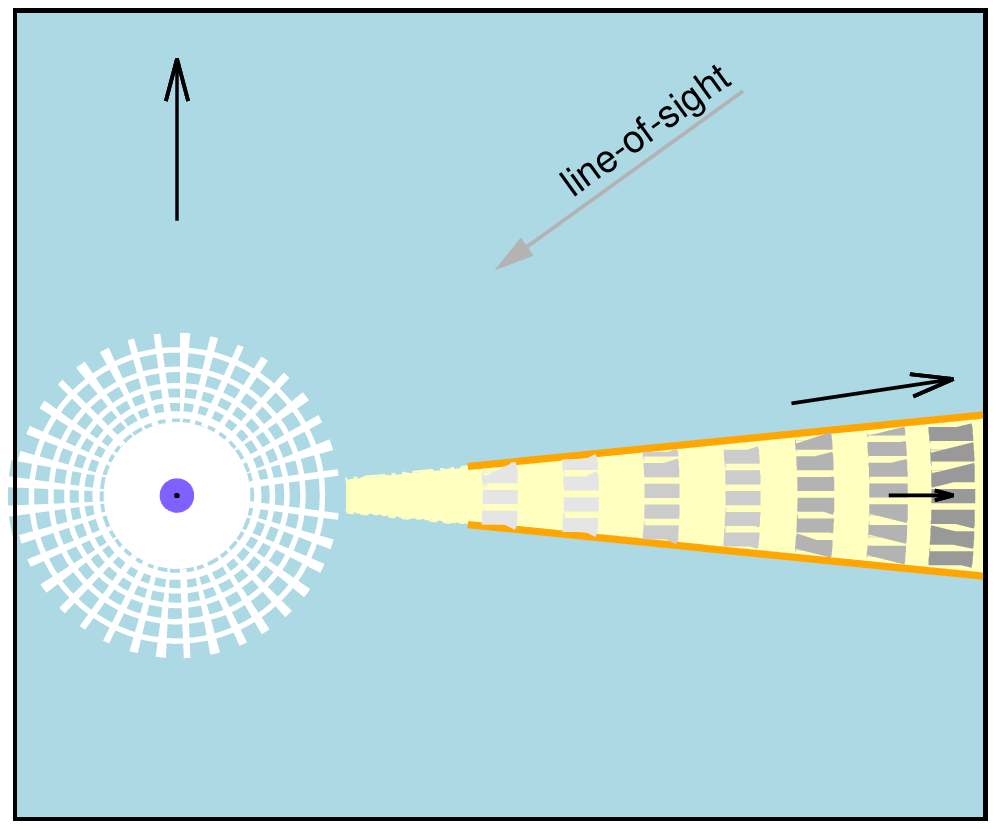}}
\end{center}
\caption{
Sketch for nova ejecta inferred from the model SED on day 35 
(top) and 100 (bottom). A slow equatorially concentrated 
outflow (yellow area) can be formed due to rotation of the WD 
\citep[][]{cask12} and/or by the binary stars interaction with 
the ejecta \citep[][]{livio+90}. A higher velocity and more 
spherical wind (light-blue) is powered by the WD. Bow shocks 
are produced between both the ejecta (orange lines), where dust 
can form (gray elements; Sect.~\ref{sss:dform}). 
Just before the steep drop in the star's brightness, the day 35 
model SED indicates an oblate shape of the WD pseudophotosphere 
(dark-blue; Sect.~\ref{sss:wd}). 
After the optical drop, the day 100 model SED indicates stopping 
the WD wind (white; Sect.~\ref{sss:mdot}) and dusty disk with 
a hotter and cooler dust (light- and dark-gray elements; 
Sects.~\ref{sss:ddisk} and \ref{sss:dform}). 
A similar structure was inferred from radio imaging of V959~Mon 
\citep[see Fig.~3 of][]{chomiuk+14}. 
}
\label{fig:sketch}
\end{figure}

The day 100 model SED indicates the simultaneous presence of 
a strong dust emission and luminous high temperature super-soft 
X-ray source in the nova. 
This fact and no detection of the dust extinction along the 
line of sight (Sect.~\ref{sss:dust}) constrain a non-spherical 
distribution of the dust within the ejecta. According to the 
biconical ionization structure of the ejecta with an equatorially 
concentrated outflow \citep[see Fig.~9 of][]{sk+14}, 
we will assume that the dust is located within the cooler and 
denser equatorial zone, where it is shielded from the hard 
radiation of the WD (see Fig.~\ref{fig:sketch}). 
Shaping of the dust into a flat disk/ring around the heating 
source is also supported by the long-lasting presence of the 
dust, to $\sim$day 680 \citep[][]{evans+17}, because its radial 
thickness for the incident radiation is extremely large 
($\approx R_{\rm D}^{\rm eff} \sim 10^{14}$\,cm; 
Table~\ref{tab:par}). 
 
From day 72, the models SED indicate two components of the 
dust emission (Fig.~\ref{fig:sedir}), whose size and radiation 
are very different. Their effective radii, $R_{\rm D}^{\rm eff}$ 
and luminosities, $L_{\rm D}$, are in Table~\ref{tab:par}. 
Approximating their geometry with a flat disk seen under the 
inclination angle $i$, their radius 
$R_{\rm disk} = R_{\rm D}^{\rm eff}/\sqrt{\cos(i)}$ and 
the luminosity $L_{\rm disk} = L_{\rm D}/\cos(i)$ 
(see Appendix~B). 
Location of the dust in the ejecta is sketched in 
Fig.~\ref{fig:sketch}. 
%
%

\subsubsection{Formation of the dust in the equatorial disk}
\label{sss:dform}
Recently, \cite{derdzinski+17} proposed that dust formation 
occurs in the cool, dense shell behind the powerful radiative 
shocks within nova outflows. Assuming that the TNR is first 
accompanied by a slow ejection of mass focused in the equatorial 
plane and then followed by a second ejection with a higher 
velocity and more spherical geometry, the subsequent collision 
between both the ejecta produces strong shocks in the equatorial 
plane. 
As the shocked gas cools by a factor of $\sim$10$^3$ and 
increases its density with a similar factor, it represents 
an ideal environment for dust forming. 

Plausibility of this interpretation for V339~Del is supported 
by the following observations. 
\begin{enumerate}
\item
The primary outflow with the expansion velocity of 
1600 -- 700\kms\ was followed by the secondary outflow 
in the form of a wind accelerating to $\sim$2700\kms\ just 
after the fireball phase 
\citep[$\gtrsim$day 6, see Figs.~7 and 8 of][]{sk+14}. 
\item
Detection of the $\gamma$-ray emission near the optical peak 
with a maximum around day 8 \citep[see][]{acker+14,ahnen+15} 
confirmed the presence of shocks developing between the fast 
and slow flows \citep[see also][ orange lines in 
Fig.~\ref{fig:sketch} here]{cheung+16,li+17}. 
\item
\cite{kawakita+19} attributed the change in the effective 
geometry of V339 Del (see Sect.~\ref{s:intro}) to the 
collision between the slowly expanding torus with the faster 
nova wind, which is consistent with the model proposed by 
\cite{li+17}. 
\item
According to \cite{h+k18}, the slow decline of the optical flux 
$F \propto t^{-1}$, observed for some novae during the early 
stage of their evolution, can be caused by the shock interaction 
that decelerates the ejecta resulting in a slower decrease of 
its density, and thus the flux with the time $t$. After the shock 
breakout, the slope changes to $F \propto t^{-1.6}$. 
Such slopes of the nova decay are similar to those observed 
for V339~Del (see Fig.~\ref{fig:slopes}), which is consistent 
with the presence of a shock deceleration mechanism. 
\end{enumerate}
The hotter, smaller and less luminous dust component may 
represent just the innermost part of the dusty equatorial 
ring (light-gray elements in Fig.~\ref{fig:sketch}, bottom). 
Progressively harder radiation of the WD heats up the dust and 
destroys it gradually toward to its outer rim. 
The temperature of the hotter dust, $\sim$1700\,K, which is just 
above the upper limit for the condensation of graphitic carbon 
($\sim$1690\,K) -- the main component of the dust in V339~Del 
\citep[see][]{evans+17}, is consistent with its destruction. 
As a result, the dust emission will gradually weaken. 

Finally, we note that the cooler, equatorially concentrated 
outflow could be formed during outburst by compression of the 
mass-outflow (wind) toward the equatorial plane due to rotation 
of the WD as was suggested for Z~And-type outbursts of symbiotic 
binaries by \cite{cask12}\footnote{Although, binary components 
orbiting within the nova envelope can also focus the mass toward 
the orbital plane \citep[][]{livio+90,lloyd+97}.}. 
This mechanism \citep[originally introduced by][]{bjorkcass93}
naturally gives rise to the biconical ionization structure 
of the ejecta, whose opening is related to the mass-loss rate 
from the WD as $\dot M_{\rm WD}^{-2}$. Thus the decrease of 
$\dot M_{\rm WD}$ (see Sect.~\ref{sss:mdot}) will cause opening 
of the ionized zone and narrowing the dense dusty disk at the 
equator, whose innermost part will move away from the central 
WD \cite[see Fig.~1 of][]{cask12}. 

\subsection{Long-lasting super-Eddington luminosity}
\label{ss:lum}
\subsubsection{The case of V339~Delphini}
\label{sss:lv339}
\begin{figure}[p!t]
\begin{center}
\resizebox{\hsize}{!}{\includegraphics[angle=-90]{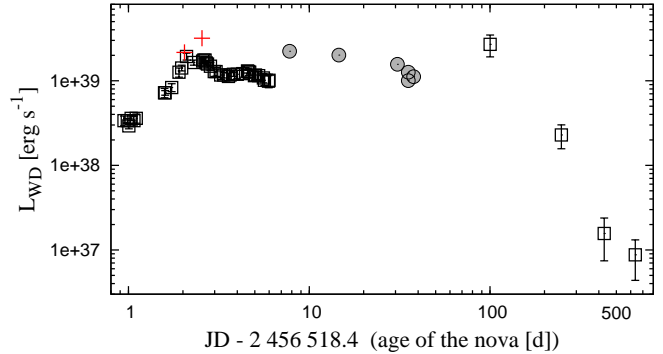}}
\end{center}
\caption{
Evolution of the WD luminosity along the nova age. 
Squares denote values derived from the stellar component 
of radiation in the models SED 
\citep[see][ for days $<$7; and this paper for days $>$99]{sk+14}, 
while circles represent those obtained from the nebular 
component (Sect.~\ref{ss:lrwd}). 
Red crosses are values published by \cite{taranova+14} and 
\cite{gehrz+15}. 
}
\label{fig:evoll}
\end{figure}

Figure~\ref{fig:evoll} shows evolution of $L_{\rm WD}$ along 
the nova age, from the first observations around day 1 to 
the last model SED on day 636. 
During the early evolution, from day $\sim$1 to day $\sim$2, 
the real increase of $L_{\rm WD}$ from $\sim 3.5\times 10^{38}$ 
to $\sim 1.8\times 10^{39}$\es\ was indicated by the increase 
of fluxes from the WD pseudophotosphere, whose maximum SED was 
located in the optical \citep[see Fig.~1 of][]{sk+14}. After 
following decrease to 
$\sim 1.2\times 10^{39}$\es\ during day $\sim$3, the $L_{\rm WD}$ 
persisted at a high level of $1-2\times 10^{39}$\es\ until 
day 100, when the WD radiation was determined by the X-ray/near-IR 
model SED (Fig.~\ref{fig:sed100}). Such the high luminosity was 
probably prolonged to day 150, when {\em Swift}-XRT still 
detected strong super-soft X-ray emission at the same level 
as on day 100. 
A significant decrease of $L_{\rm WD}$ as well as $T_{\rm BB}$ 
was first determined on day 248 (Appendix~\ref{s:appA}) 
as indicated by weakening of the X-ray source and the decrease 
of the UV fluxes. Following measurements on day 428 and 636 
confirmed gradual fading of $L_{\rm WD}$ to $\sim 10^{37}$\es\ 
(Sect.~\ref{ss:lrwd}). 

Using the indirect method in deriving $L_{\rm WD}$, the 
principal assumption is that the nebula is optically thick for 
the Lyman continuum photons of the ionizing source. Before 
the SSS phase, observations support this assumption 
(see Sect.~\ref{ss:lrwd}). Therefore, we estimated $L_{\rm WD}$ 
also for \textsl{EM} determined by \cite{sk+14} for 
day 7.8 to 37.8 (see their Table~3). We adopted 
$T_{\rm BB} \equiv 73000$\,K, at which the ionizing source 
produces maximum of photons capable of ionizing hydrogen 
for a given luminosity. 

At the end of the fireball stage ($\sim$day 6), values of 
$L_{\rm WD}$ from the WD pseudophotosphere were lower by the 
fraction generating the nebular radiation that started to be 
visible at this time \citep[see Figs.~1 and 2 of][]{sk+14}. 
Because the following values of $L_{\rm WD}$ from the \textsl{EM} 
could be overestimated due to collisional ionization, but also 
underestimated due to measuring only the optically thin part 
of the nebula, it is not possible to determine errors for 
indirectly estimated values of $L_{\rm WD}$. 

Nevertheless, $L_{\rm WD}$ given by the stellar component of 
radiation in the day 100 model SED confirms previous values 
from \textsl{EM}, and thus the long-lasting super-Eddington 
luminosity for, at least, the first 100 days of the nova life 
(Fig.~\ref{fig:evoll}). 
%
%
%

\subsubsection{When the super-Eddington luminosity 
               can be indicated?}
\label{sss:isuper}
In most cases the super-Eddington luminosity is indicated at 
the optical maximum, when the absolute magnitude of numerous 
(fast) novae, $M_{\rm V,B} \lesssim -7$ 
\citep[e.g.][]{livio92,valle+95,shafter+09,shav+dot12}. 
This is because during this very early stage of evolution, most 
novae reprocess significant fraction of their radiation into 
the optical, which allows a direct estimate of their luminosity. 
Usually, a simple blackbody fit to the multicolor photometry 
provides a rough estimate of the nova luminosity 
\citep[e.g.,][ for V339~Del]{gehrz+15}. 

Later, when the maximum SED is shifting to shorter 
wavelengths and the optical is dominated by the nebular 
emission, it is not possible to determine directly the nova 
luminosity by fitting only the observed UV--optical continuum 
(e.g., the day 35 model SED here). This possibility arises 
again during the SSS phase, when the super-soft X-ray and 
far-UV fluxes define the WD spectrum from both the short- and 
long-wavelength side (here, the day 100 model SED). 
This can reveal a startling result that the super-Eddington 
state of some novae lasts for a long time after their 
eruption (Sect.~\ref{sss:lv339}). 
To prove this case, at least for the optically bright novae, 
measurements of X-ray and far-UV fluxes during the SSS phase 
are required. 

\subsubsection{Current view on super-Eddington luminosity}
\label{sss:selum}
Our findings of the super-Eddington luminosity for V339~Del is 
not consistent with theoretical modeling that the super-Eddington 
phase can persist only for a short time at the beginning of the 
nova eruption 
\citep[e.g.,][]{prial+kov95,yaron+05,starrfield+08}. 
On the other hand, the long-term super-Eddington luminosity was 
observationally documented for more novae, for example, nova 
FH~Ser \citep[][]{friedjung87}, LMC~1988 $\#$1 \citep[][]{schwarz+98}, 
LMC~1991 \citep[][]{schwarz+01}, RS~Oph \citep[][]{sk15b,sk15c} 
and nova SMCN~2016-10a \citep[][]{aydi+18}, which thus support 
our finding for V339~Del. 
Also, the summary of \cite{shafter+09} that there are 12, 26 and 5 
novae in our Galaxy, M31 and LMC, respectively, which have reached 
absolute visual magnitude $M_{\rm V} \lesssim -9.0$, suggests 
luminosities highly above the Eddington limit. 

The super-Eddington state of novae was investigated by 
\cite{shaviv98}, who suggested a decrease of the effective 
opacity of the inhomogeneous atmosphere of novae, which increases 
the Eddington luminosity well above its standard value, calculated 
for the Thomson-scattering opacity 
\citep[see also][]{shav+dot10,shav+dot12}. 

Based on remarkable correlation between the $\gamma$-ray and 
optical LCs of the luminous ($\approx 10^{39}$\es) nova 
ASASSN-16ma, \cite{li+17} suggested that the majority 
of the optical light comes from reprocessed emission from 
shocks rather than the WD. 
The $\gamma$-ray--optical brightness correlation was indicated 
at/after the optical maximum (see their Fig.~1), when the dense 
slow ejecta ahead of the shocks reprocesses their radiation 
into the optical, giving rise to the observed optical luminosity 
of $\approx 10^{39}$\es. 
In this way, the authors replace the standard model, in which 
most of ultraviolet and/or optical emission in novae is the 
result of outwards diffusion of the radiation from the WD 
\citep[e.g.,][]{yaron+05}, and state that the shock-driven 
emission provides an explanation for super-Eddington luminosity 
observed for many novae. 
However, at late stages, particularly during the SSS phase, 
the super-Eddington luminosity is indicated by direct measuring 
the photospheric radiation transferred through the atmosphere 
of the WD, which thus leaves the long-standing mystery of why 
many novae exceed the Eddington limit for a long time still open. 

\subsubsection{On the long-lasting super-Eddington luminosity}
\label{sss:long}
A possibility how to keep a nova at the super-Eddington 
state for a long time is to fuel the burning WD also after 
the eruption. 
Here, by analogy with the Z~And-type outbursts, a disk-like 
reservoir of mass can be created at the equatorial plane during 
the outburst (see the last paragraph of Sect.~\ref{sss:dform}). 
Due to the presence of a strong central source of radiation 
(the burning WD), the inner parts of the disk can be accreted 
again onto the WD via the radiation-induced warping, which 
prolongs the period with a high luminosity, until depletion 
of the inner disk. This event can be accompanied by formation 
of bipolarly collimated high-velocity outflow 
\citep[see][ and references therein]{sk+18}. The long-term 
super-Eddington luminosity of the recurrent nova RS~Oph 
\citep[][]{sk15b,sk15c} and the bipolar jet-like collimated 
outflow observed in its spectrum during day 10 to 30 after 
the 2006 outburst maximum \citep[see][]{sk+08} suggest that 
this mechanism could work also during outbursts of classical 
novae. 
However, multiwavelength observations along evolution of other 
novae and their theoretical modeling are needed to justify 
applicability of this accretion mechanism for classical novae. 

\subsection{Measuring $N_{\rm H}$ from Rayleigh scattering}
\label{ss:ray}
In some recent papers on novae the authors used the Voigt 
function to model the interstellar Ly-$\alpha$ absorption 
\citep[e.g.][]{mason+18}. 
Instead, we modeled the attenuation around the Ly-$\alpha$ 
line by Rayleigh scattering on atomic hydrogen. 
We justify our approach as follows. 

Rayleigh scattering by neutral atoms of hydrogen represents 
the process, where an incident photon raises an electron from 
the ground state to the intermediate state, followed by the 
direct return of the electron to the original state, re-emitting 
a photon of the same energy \citep[e.g.,][]{nussb+89}. 
The strength of Rayleigh scattering is determined by the value 
of $N_{\rm H}$ and its profile is given solely by its cross-section 
(Sect.~\ref{sss:sed35}). As the cross-section has a `singularity' 
near the wavelength of a Lyman line 
\citep[e.g., Fig~2 of][]{nussb+89}, the rest flux around the 
reference wavelength is close to zero. 
For example, $N_{\rm H} = 1\times10^{21}$\cmd\ creates an 
absorption core with the zero rest flux\footnote{Contribution 
          of the scattered photons to the 
          line-of-sight, which can be of a few percents 
          \citep[][]{schmid95}, is not detectable on our 
          spectra.}, 
the FWHM of $\sim 15$\,\AA\ and wings expanding approximately 
to $\pm 25$\,\AA\ around the Ly-$\alpha$ line 
(see Fig.~\ref{fig:rayall}). 
Therefore, in spite of small kinematic motions of atoms in 
the ISM, Rayleigh scattering can create the very broad hollow 
with expanding wings around the Ly-$\alpha$ line, depending 
only on the quantity of $N_{\rm H}$. 
 
\section{Summary}
\label{s:sum}
In this paper we continued the work of \cite{sk+14} by 
multiwavelength modeling the SED of V339~Del from the 
iron-curtain phase on day 35 to the nebular phase on day 636 
(Sect.~\ref{s:analysis}, Figs.~\ref{fig:sed35}, \ref{fig:sed100}, 
\ref{fig:seduvop} and \ref{fig:sedir}). 
Pivotal models were made for day 35 and 100, when the large 
and most important part of the nova spectrum is covered by 
simultaneous observations (Table~\ref{tab:obs}). 
In this way we determined physical parameters of the stellar, 
nebular and dust component of radiation (Table~\ref{tab:par}) 
and obtained new information on the nova evolution. The main 
results can be summarized as follows. 
\begin{enumerate}
\item
During the transition from the iron-curtain to the SSS phase 
(days 35 -- 72), the steep decline in the optical brightness 
by 3--4\,mag, the fall of $N_{\rm H}$ by its CSM component 
from $\sim 10^{23}$ to $\sim 10^{21}$\cmd\ and the decrease 
of \textsl{EM} from $\sim 2\times 10^{62}$ to 
$\sim 8.5\times 10^{60}$\cmt\ (Fig.~\ref{fig:nhem}), were 
caused by stopping-down the wind from the WD 
(Sect.~\ref{ss:transition}). 
\item
Models SED on day 35 and 100 revealed the long-lasting 
super-Eddington luminosity that persists at the level of 
$1-2\times 10^{39}\,(d/4.5\kpc)^2$\es\ from $\sim$day 2 
(the optical maximum) to at least day 100 of the nova life 
(Sect.~\ref{ss:lum}, Fig.~\ref{fig:evoll}). 
%
%
\item
The day 35 model SED indicates an oblate shape of the WD 
pseudophotosphere, which hotter part with a smaller radius is 
located around the poles, while the cooler regions with larger 
radii are stretched towards the WD equator (Sect.~\ref{sss:wd}). 
The indicated dust can be formed within a slow equatorially 
concentrated outflow as a result of its interaction with the 
fast nova wind (Sect.~\ref{sss:dform}, Fig.~\ref{fig:sketch}). 
The dust emission reached its maximum around day 59 
(Figs.~\ref{fig:slopes} and \ref{fig:sedir}). 

On day 100, the co-existence of both the strong dust emission 
and the luminous high temperature WD photosphere confirms the 
disk-like shaping of the dust even during the SSS phase 
(Sect.~\ref{sss:ddisk}). 
The indicated hotter dust is located at the inner part of 
the dusty ring, where the hard WD radiation destroys the dust 
gradually toward to its outer rim. In this way the disk can 
preserve the dust within the ejecta for a long time 
(Sect.~\ref{sss:dform}, Fig.~\ref{fig:sketch}). 
\end{enumerate}
%
%

\acknowledgments
The author thanks the anonymous referee for constructive 
suggestions that led to improving the manuscript. 
\textsl{HST} spectra presented in this paper were obtained from
the Mikulski Archive for Space Telescopes (MAST). MAST is 
located at the Space Telescope Science Institute (STScI). 
STScI is operated by the Association of Universities for  
Research in Astronomy, Inc., under NASA contract NAS5-26555.
This work is in part based on observations obtained with 
\textsl{XMM-Newton}, an ESA science mission with instruments 
and contributions directly funded by ESA Member States and NASA. 
Mitsugu Fujii, Taya Tarasova and David Boyd are thanked for 
acquisition of their spectra at the Fujii Kurosaki Observatory, 
the Crimean Astrophysical Observatory and the West Challow 
Observatory, respective. 
We also acknowledge the variable-star observations from the AAVSO 
International Database contributed by observers worldwide and used 
in this research. 
This work was supported by the Slovak Research and Development 
Agency under the contract No. APVV-15-0458, by the Slovak 
Academy of Sciences grant VEGA No. 2/0008/17 and by the 
realization of the project ITMS No.~26220120029, based on 
the supporting operational Research and development program 
financed from the European Regional Development Fund.
\appendix
\label{app}
\section{Temperature of the WD pseudophotosphere on day 248}
\label{s:appA}
\begin{figure}
\begin{center}
%
\resizebox{15cm}{!}{\includegraphics[angle=-90]{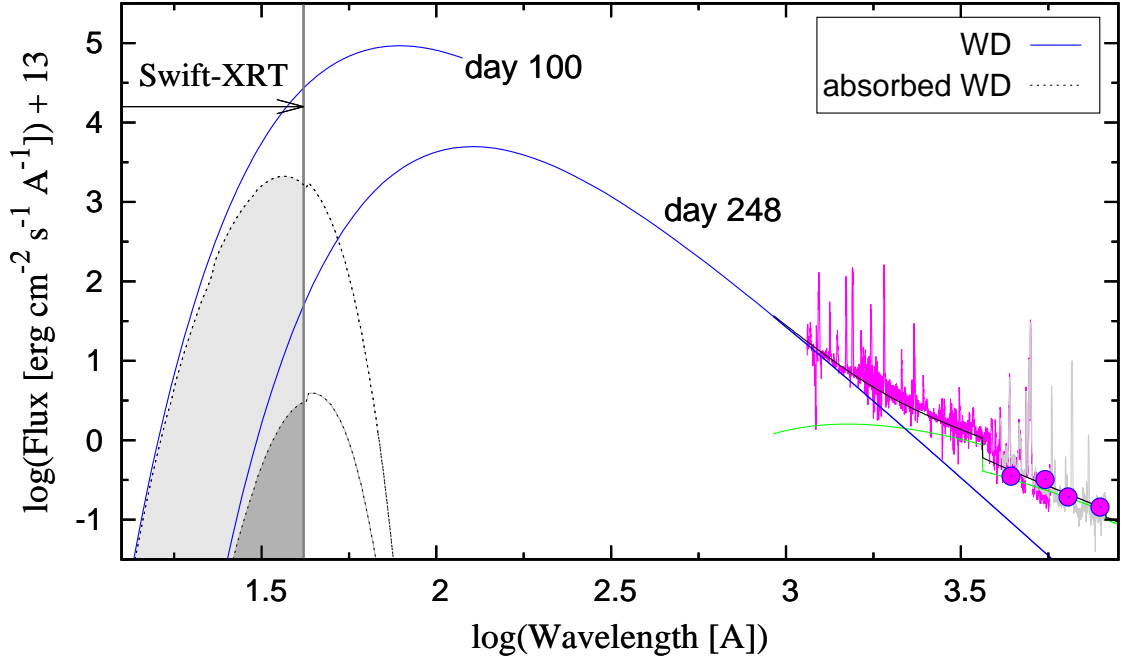}}
\end{center}
\caption{
On day 248, the WD radiation was determined by scaling its 
long-wavelength part to dereddened far-UV fluxes and the 
temperature, at which the radiation produces the X-ray photons 
rate as measured by {\em Swift-}XRT (i.e., above 0.3\,keV; 
the gray area). Compared is the absorbed and model WD radiation 
from day 100 (see text). 
}
\label{fig:t248}
\end{figure}

To determine the WD temperature on day 248 we assume that both 
the far-UV fluxes and the super-soft X-ray photons are produced 
by the long- and short-wavelength part of the WD radiation. 
The former is given by the model SED (see Fig.~\ref{fig:t248}) 
and the latter by the super-soft X-ray source as suggested by 
the low hardness ratio 
F(0.45--1\,keV)/F(0.3--0.45\,keV)$\sim$0.1 on day 248 
\citep[see Fig.~1 of][]{shore+16} and no X-ray flux 
detection below 22\,\AA\ during the maximum of the SSS phase 
\citep[][]{nelson+13,ness+13}. 
Under this assumption we can write the ratio 
\begin{equation}
 \frac{Q^{100}(XRT)}{Q^{248}(XRT)} =
 \frac{c^{100}(XRT)}{c^{248}(XRT)} = R, 
\label{eq:qratio}
\end{equation}
where $Q^{100}(XRT)$ and $Q^{248}(XRT)$ are photon rates 
(s$^{-1}$) emitted by the WD photosphere within the 
{\em Swift}-XRT range (0.3--10\,keV) attenuated with 
$N_{\rm H}$, while c$^{100}(XRT)$ and c$^{248}(XRT)$ are photon 
rates directly measured by {\em Swift}-XRT on day 100 and 248, 
respectively. 
According to Eq.~(11) of \cite{sk01}, the flux of quanta 
$Q(XRT)$ can be expressed as 
\begin{equation}
Q(XRT) = \frac{L_{\rm WD}}{\sigma T_{\rm BB}^{4}}
         f^{\rm abs}(T_{\rm BB}), 
\label{eq:Q}
\end{equation}
where the function
\begin{equation}
f^{\rm abs}(T_{\rm BB}) = \frac{\pi}{hc}\int_{XRT}\!\!\!\lambda\, 
    B_{\lambda}(T_{\rm BB})\,
    e^{-\sigma_{\rm X}(\lambda)\,N_{\rm H}}\,\rm d\lambda
\label{eq:fta}
\end{equation}
represents the flux of photons emitted by 1\,cm$^2$ area of 
the WD photosphere within the {\em Swift}-XRT range, absorbed 
by hydrogen column density $N_{\rm H}$. 
Using  Eqs.~(\ref{eq:Q}) and (\ref{eq:qratio}), the luminosity 
of the WD on day 248 can be written as 
\begin{equation}
 L^{248}_{\rm WD} = \frac{Q^{100}(XRT)}{R}
   \frac{\sigma (T_{\rm BB}^{248})^4}{f^{\rm abs}(T_{\rm BB}^{248})}, 
\label{eq:l248}
\end{equation}
which for $L_{\rm WD} = 
4\pi d^2 \theta_{\rm WD}^2 \sigma T_{\rm BB}^4$ 
and the scaling factor $\theta_{\rm WD}^2$ = 
$F_{\rm WD}(\lambda)/\pi B_{\lambda}(T_{\rm BB})$
\citep[e.g., Eqs.~(5) and (6) of][]{sk05} 
provides equation, 
\begin{equation}
 \frac{q^{100}(XRT)}{R}
 \frac{\pi B_{\lambda}(T_{\rm BB}^{248})}{F_{\rm WD}^{248}(\lambda)} - 
 f^{\rm abs}(T_{\rm BB}^{248}) = 0, 
\label{eq:solution}
\end{equation}
where $q^{100}(XRT) = Q^{100}(XRT)/4\pi d^2$ (cm$^{-2}$\,s$^{-1}$). 
For the parameter $q^{100}(XRT)$ given by Eq.~(\ref{eq:Q}), $R$ 
derived from observations and the flux of the WD, 
$F_{\rm WD}^{248}(\lambda)$, 
the solution of Eq.~(\ref{eq:solution}) provides the WD 
temperature and Eq.~(\ref{eq:l248}) its luminosity on day 248. 

For the measured photon rates c$^{100}(XRT)$ = 35$\pm$5 
and c$^{248}(XRT)$ = 0.025$\pm$0.005\,s$^{-1}$ 
\citep[see Fig.~1 of][]{shore+16}, i.e., $R = 1500\pm 500$, 
$Q^{100}(XRT) = 1.1\times 10^{46}$\,s$^{-1}$ (for $L_{\rm WD}$, 
$T_{\rm BB}$ and $N_{\rm H}$ in Table~\ref{tab:par}), i.e., 
$q^{100}(XRT) = 4.54$\,cm$^{-2}$\,s$^{-1}$ and the WD flux 
at, e.g., $\lambda = 1195$\,\AA, 
$F_{\rm WD}^{248}(1195) = 1.37\times 10^{-12}$\ecsa, 
Eq.~(\ref{eq:solution}) and Eq.~(\ref{eq:l248}) yield 
$T_{\rm BB}^{248} = 227\pm 5$\,kK and 
$L^{248}_{\rm WD} = (2.3\pm 0.6)\times 10^{38}$\es, 
respectively. 
Figure~\ref{fig:t248} demonstrates principle of this approach 
in a graphical form. 
%
%
%
%
%
\section{Dust emission from a flat disk}
\label{s:appB}
According to Sect.~\ref{sss:ddisk} the geometry of emitting 
dust can be approximated by a flat disk encompassing the central 
heating source. If the disk is optically thick and radiates 
locally like a blackbody, the observed flux distribution of the 
disk, $F_{\lambda}$, at a distance $d$ is given by contributions 
from blackbody annuli, 2$\pi\,r\,dr$, integrated through 
the entire disk, i.e. 
\begin{equation}
 F_{\lambda} = \frac{2\pi\cos(i)}{d^{2}}
               \int_{R_{\rm in}}^{R_{\rm disk}}\! 
               B_{\lambda}(T_{\rm disk}(r))\,r\,{\rm d}r,
\end{equation}
where $i$ is the angle between the line of sight and the normal 
to the disk plane, $R_{\rm disk}$ its radius and $T_{\rm disk}(r)$ 
the radial temperature structure of the disk. 
According to the model SED, $T_{\rm disk}(r)$ can be assumed 
to be constant throughout the dusty disk. Thus, assuming 
$T_{\rm disk}(r) = T_{\rm D}$ and the inner radius 
$R_{\rm in}\ll R_{\rm disk}$, the observed bolometric flux of 
the disk can be approximated by 
\begin{equation}
 F_{\rm bol} = \left(\frac{R_{\rm disk}}{d}\right)^2 \cos(i)\,
               \sigma T_{\rm D}^4 ,
\end{equation}
which gives the disk radius, 
\begin{equation}
 R_{\rm disk} = \left(\frac{d^2 F_{\rm bol}}
             {\sigma T_{\rm D}^4 \cos(i)}\right)^{1/2} 
              = R_{\rm D}^{\rm eff}/\sqrt{\cos(i)}
\end{equation}
and the disk luminosity, 
\begin{equation}
 L_{\rm disk} = 4\pi d^2 F_{\rm bol}/\cos(i) 
              = L_{\rm D}/\cos(i)
\end{equation}
%
%
%

%
\end{document}